\documentclass[a4paper,11pt]{article}
\pdfoutput=1 %
\usepackage{jcappub} 

\usepackage[T1]{fontenc} 

\usepackage[utf8]{inputenc}

\usepackage{hyperref} 
\usepackage{url}
\usepackage{graphicx}
\usepackage{esvect}
\usepackage{graphics}
\usepackage{natbib}
\usepackage{mathrsfs}
\usepackage{blindtext}

\usepackage{comment}

\usepackage{lineno}

\usepackage[table]{xcolor}
\usepackage{multirow}

\usepackage[normalem]{ulem}
\usepackage{color}

\usepackage{soul}

\usepackage{supertabular}
\usepackage{tabularx}

\title{Galactic Gas Models Strongly Affect the Determination of the Diffusive Halo Height}

\author[a, b, c]{Pedro De La Torre Luque}
\emailAdd{pedro.delatorre@uam.es}
\author[c]{Tim Linden}
\emailAdd{linden@fysik.su.se}

\affiliation[a]{Departamento de F\'{i}sica Te\'{o}rica, M-15, Universidad Aut\'{o}noma de Madrid, E-28049 Madrid, Spain}
\affiliation[b]{Instituto de F\'{i}sica Te\'{o}rica UAM-CSIC, Universidad Aut\'{o}noma de Madrid, C/ Nicol\'{a}s Cabrera, 13-15, 28049 Madrid, Spain}
\affiliation[c]{The Oskar Klein Centre, Department of Physics, Stockholm University, AlbaNova\\ SE-10691 Stockholm, Sweden}
\date{\today}

\abstract{The height of the Milky Way diffusion halo, above which cosmic-rays can freely escape the galaxy, is among the most critical, yet poorly known, parameters in cosmic-ray physics. Measurements of radioactive secondaries, such as $^{10}$Be or $^{26}$Al, which decay equivalently throughout the diffusive volume, are expected to provide the strongest constraints. This has motivated significant observational work to constrain their isotopic ratios, along with theoretical work to constrain the cross-section uncertainties that are thought to dominate radioactive secondary fluxes. In this work, we show that the imprecise modelling of the Milky Way spiral arms significantly affects our ability to translate $^{10}$Be and $^{26}$Al fluxes into constraints on the diffusive halo height, biasing our current results. Utilizing state-of-the-art spiral arms models we produce new predictions for the $^{10}$Be and $^{26}$Al fluxes that motivate upcoming measurements by AMS-02 and HELIX.}

\begin{document}
\maketitle
\flushbottom

\section{Introduction}
\label{sec:intro}

The recent explosion in the quality and quantity of cosmic-ray (CR) measurements has greatly improved our understanding of CR acceleration and transport throughout the Milky Way. Understanding CR physics can, in turn, provide valuable methods for constraining Galactic magnetic fields~\cite{Jansson_2012_2}, astrophysical plasmas~\cite{Fornieri:2020wrr}, Milky Way gas~\cite{Widmark:2022qgx}, diffuse non-thermal emission~\cite{2012ApJ...750....3A} and galaxy evolution~\cite{GalFormation, Hopkins, CR-Galaxy, Farber_2018}. Moreover, a detailed understanding of astrophysical CRs allows us to set strong constraints on novel physics such as dark matter annihilation, both through the direct observation of CR antiparticles~\cite{Giesen:2015ufa, Cuoco:2016eej, Cui:2016ppb}, as well as through indirect observations at $\gamma$-ray~\cite{Daylan:2014rsa, Tibaldo:2015ooa, McDaniel:2023bju} and radio~\cite{Wang:2023sxr} energies.


To model the CR flux at Earth, physicists have developed detailed computational models of galactic CR production (primarily in supernovae remnants and pulsars), propagation (primarily diffusion through plasma turbulence~\cite{Berezinskii, PTUSKIN201244, Blasi_2012, Fornieri_2021}), and interactions (including radiative losses, hadronic interactions, diffusive escape, and spallation~\cite{Berezinskii, DRAGON2-1, Strong_1998, Strong_2001}. Notably, several of these energy loss-mechanisms (radioactive decay, hadronic interactions and spallation) produce new ``secondary" CRs that include new particle (and anti-particle) content, which subsequently also propagate through the Milky Way. By measuring the plethora of available signals, along with their spectral information, we can study CR physics. Of particular interest are the calculated ratios of primary and secondary CR species, which isolate the particular physics of secondary production and diffusion.

However, CR models include many degeneracies that make it difficult to translate CR measurements into theoretical understanding. Foremost among these is the well-known degeneracy between the height of the galactic diffusive halo (H) and the diffusion coefficient throughout the bulk of the Milky Way (D). This degeneracy stems from the fact that the Boron-to-Carbon (B/C) ratio, which is the best measured primary-to-secondary ratio, is most sensitive to the ratio H/D~\cite{evoli2023phenomenological}. Our lack of knowledge regarding the halo height plagues not only our ability to model CRs, but has a particularly large effect on our ability to constrain dark matter (DM) models, because the DM induced CR flux extends to larger heights above the galactic plane than astrophysical sources. In CR searches, uncertainties in the halo height translate into factor of $>4$ uncertainties in the upper limits on the DM annihilation or decay rates~\cite{Genolini_2021}, while in $\gamma$-ray searches, uncertainties high-latitude astrophysical background induce significant uncertainties in dwarf spheroidal galaxy (dSPh) searches~\cite{Linden:2019soa, McDaniel:2023bju}.

Several probes have aimed to break this degeneracy and independently determine the height of the magnetized halo, including radio observations of Galactic~\cite{bringmann2012radio, OrlandoStrong2013, Di_Bernardo_2013} and extragalactic~\cite{Beck_2015} synchrotron radiation, as well as studies of diffuse X-ray and $\gamma$-ray~\cite{biswas2018constraining, Ackermann, Tibaldo_2015} emission. However, the most promising approach focuses on the detection of radioactive CR secondaries such as $^{10}$Be or $^{26}$Al. These CR secondaries are produced via CR collisions with gas in the galactic plane in a known abundance ratio compared to their stable cousins, $^{9}$Be and $^{27}$Al, and then decay in known timescales throughout the entirety of the galactic halo. Thus, the measured isotropic ratios of $^{10}$Be/$^{9}$Be and $^{26}$Al/$^{27}$Al provide a unique lever arm on the galactic halo height that is independent of the B/C ratio. The proposal of using isotropic ratios as ``cosmic-ray clocks" depends on the similarity of their $\sim$Myr decay timescales and the $\sim$Myr timescales for diffusive escape from the Milky Way~\cite{Webber1973, MuñozBe10, moskalenko2000diffuse}.

Unfortunately, isotopic identification is a difficult task, which considerably limits existing data. Present observations have large uncertainties and are typically limited to low energies (between $50$~MeV and $\sim1$~GeV). This region is also complicated from a modeling perspective for three reasons: (1) low-energy particles are highly affected by Solar modulation~\cite{1968ApJ...154.1011G}, which is difficult to model, (2) production cross-sections for B, Be and Li have large uncertainties at energies near $100$~MeV, where resonances exist and different cross-section parameterisations differ in their predicted energy dependence~\cite{thesis, Luque:2021joz, Luque:2021nxb, delaTorreLuque:2022vhm, Weinrich:2020cmw, Korsmeier:2021brc, GONDOLO2014175, GenoliniRanking, Tomassetti:2015nha, Maurin_Be10}, and (3) the diffusion coefficient at these energies is not well known due to the lack of accurate secondary-to-primary ratios below $\sim0.5$~GeV.  Fortunately, the study of secondary isotopic flux ratios significantly mitigates these sources of uncertainty. While at present the predicted ratios of $^{10}$Be still only allow us to constrain the diffusion height to be between $\sim3$~and $12$~kpc~\cite{Korsmeier:2021bkw, thesis}, near-future CR and cross-section data would be expected to significantly improve these constraints.

Here, we show for the first time that the modeling of the galactic gas structure significantly affects our predictions for the $^{10}$Be and $^{26}$Al fluxes. Because the local gas contribution strongly affects only the radioactive isotope (which cannot travel far from its production site at low energies), these uncertainties are not easily corrected by studying isotopic ratios. This causes the modeling of local gas to dominate the systematic uncertainties in the $^{10}$Be/$^{9}$Be and $^{26}$Al/$^{27}$Al flux ratios, and systematically shifts the value of the diffusive halo height that is calculated from isotopic measurements. Our results indicate that utilizing a 2D cylindrical gas distribution, as is standard in many treatments, produces 10--40\% errors in the isotopic flux ratios below $1$~GeV, which then translate into factor of up to two changes in the best-fit halo height. To correct this bias, we study 3D gas distributions that include spiral arms near the Solar position, finding that well-motivated 3D gas models can shift the best-fit isotopic ratios either up or down, depending primarily on the proximty of the Earth to the nearest spiral arm. Combining these models with state-of-the-art models for solar modulation, cross-section uncertainties, and galactic diffusion, we produce novel predictions for isotopic flux ratios that will be testable by future AMS-02~\cite{AGUILAR20211, AMS02_NaAlN} and HELIX~\cite{HELIX} data.

This paper is organized as follows. We discuss our methodology and setup in Sec.~\ref{sec:Gas_Be10}. We show the implications of adopting a spiral arm structure for the gas density for unstable secondary CR ratios in Section~\ref{sec:results}. In Sec.~\ref{sec:LocalBubble} we show the effect of including an underdense local bubble around the Solar System. We discuss other uncertainties in the predictions of unstable CR fluxes in Section~\ref{sec:OtherUncerts} and summarize our main findings in Sec.~\ref{sec:conc}.

\section{Methodology and gas distribution} 
\label{sec:Gas_Be10}

Because $^{10}$Be is radioactive, it should not travel far from its birth location (compared to stable isotopes). This is particularly true below 1~GeV, where the diffusion coefficient is low and relativistic effects do not significantly boost its half-life. Thus, the isotopic flux of $^{10}$Be measured at Earth will strongly depend on the local gas distribution. 

Throughout this work, we calculate the CR fluxes at Earth from a fit to the most recent AMS-02 data~\cite{Aguilar:2015ooa, Aguilar:2020ohx, Aguilar:2021tla, Aguilar:2018keu, AMS02_NaAlN, AGUILAR20211}, for different distributions of galactic gas. We numerically solve CR particle propagation using the {\it DRAGON2} code~\cite{DRAGON2-1, DRAGON2-2}, which includes energy losses, reacceleration, advection, production of secondary particles and decay. The best-fit propagation parameters for each model are determined through analyses of secondary-to-primary CR ratios following the same strategy as in Refs.~\cite{Luque:2021nxb, delaTorreLuque:2022vhm} and is parametrized as:
\begin{equation}
 D = D_0 \beta^{\eta}\left(\frac{R}{R_0} \right)^{\delta} .
\label{eq:DiffCoeff}
\end{equation}
Note that for the energies of interest in this work (below $\sim10$~GeV), the possible break in the diffusion coefficient above $\sim300$~GeV~\cite{genolini2017indications} is not relevant. 

The source term for secondary CR production from the spallation of CRs on the interstellar gas has the following form~\cite{maurin2002galactic, Kappl:2014hha}:
\begin{equation}
Q_{CR+ISM\rightarrow Be} (E_{Be}, \Vec{x}) = \sum_{CR }4 \pi  \,  n_{ISM} (\Vec{x}) \int^{\infty}_0 dE  \,  \phi_{CR}(E, \Vec{x})\frac{d\sigma_{CR+ISM\rightarrow Be}}{dE}(E, E_{Be}) \,\, {,}
\label{eq:sec_Sourceterm}
\end{equation}
where $\Vec{x}$ indicates the spatial dependence of each term, $n_{ISM}$ represents the interstellar gas density (assumed to be H and He, in a  9:1 ratio), $\phi_{CR}$ is the flux of CR species with masses higher than that of $^{10}$Be and $\frac{d\sigma}{dE}$ are the cross sections for $^{10}$Be production, for which the head-on approximation is usually followed. Here, we adopt the default {\it DRAGON2} cross sections~\cite{Evoli:2019wwu, Luque_MCMC, Luque:2021joz}. The CRs that produce Be are mainly the primary CRs C, O, N, Ne, Mg and Si, with a small contribution from Fe and B~\cite{GenoliniRanking}.



We consider and compare both 2D and 3D models for the galactic gas density. For the 2D case, we adopt the popular Galprop 2D~\cite{Moskalenko_2002} cylindrically symmetric gas model. This model parametrizes the density of gas as a function of the vertical radial distance to the Galactic center.
For our benchmark 3D model, we adopt the Steiman-Cameron et al.~\cite{Steiman-Cameron_2010} gas distribution model with four spiral arms. We note that this model has been extensively adopted in similar CR studies~\cite{Steppa_2020, Evoli_2021, DRAGON2-1}. 
For comparison, we also implement the Wainscoat distribution~\cite{Wainscoat}. An important difference between these models is the relative position of the Earth with respect to the spiral arms, with the Wainscoat distribution setting the Solar System farther from the closest Carina-Sagittarius arm (see: Fig.~\ref{fig:Be10Maps_App} of App.~\ref{sec:AppendixA}). 

In DRAGON2 the width of the arms is regulated by a Gaussian function around the center of the arm with a standard deviation, $\sigma$ (the width scale of the arms) that we can adjust. This distribution is convoluted with the (radial and vertical) distribution of the interstellar gas, changing its azimuthal distribution to follow the arm-like pattern that we observe in the Galaxy. Importantly, we note that the total gas density in our 3D spiral arms model is such that the total gas mass is equivalent to the 2D model, which removes any effects from the overall gas normalization. The arm distribution is also implemented for the interstellar radiation field energy density and the distribution of sources (see also Ref.~\cite{Gaggero:2013eaa, Gaggero3D}), which do not affect the production of secondary particles, nor the ratios of secondary CRs. More details about the implementation of this distribution can be found in Ref.~\cite{DRAGON2-1}. We show the results for alternative gas density maps, with different values of $\sigma$ in Fig.~\ref{fig:Be10Maps_App}, in Appendix~\ref{sec:AppendixA}.

\section{Results: Impact of the gas distribution on radioactive CRs}
\label{sec:results}

\begin{figure}[!t]
\centering
\includegraphics[width=0.496\textwidth] {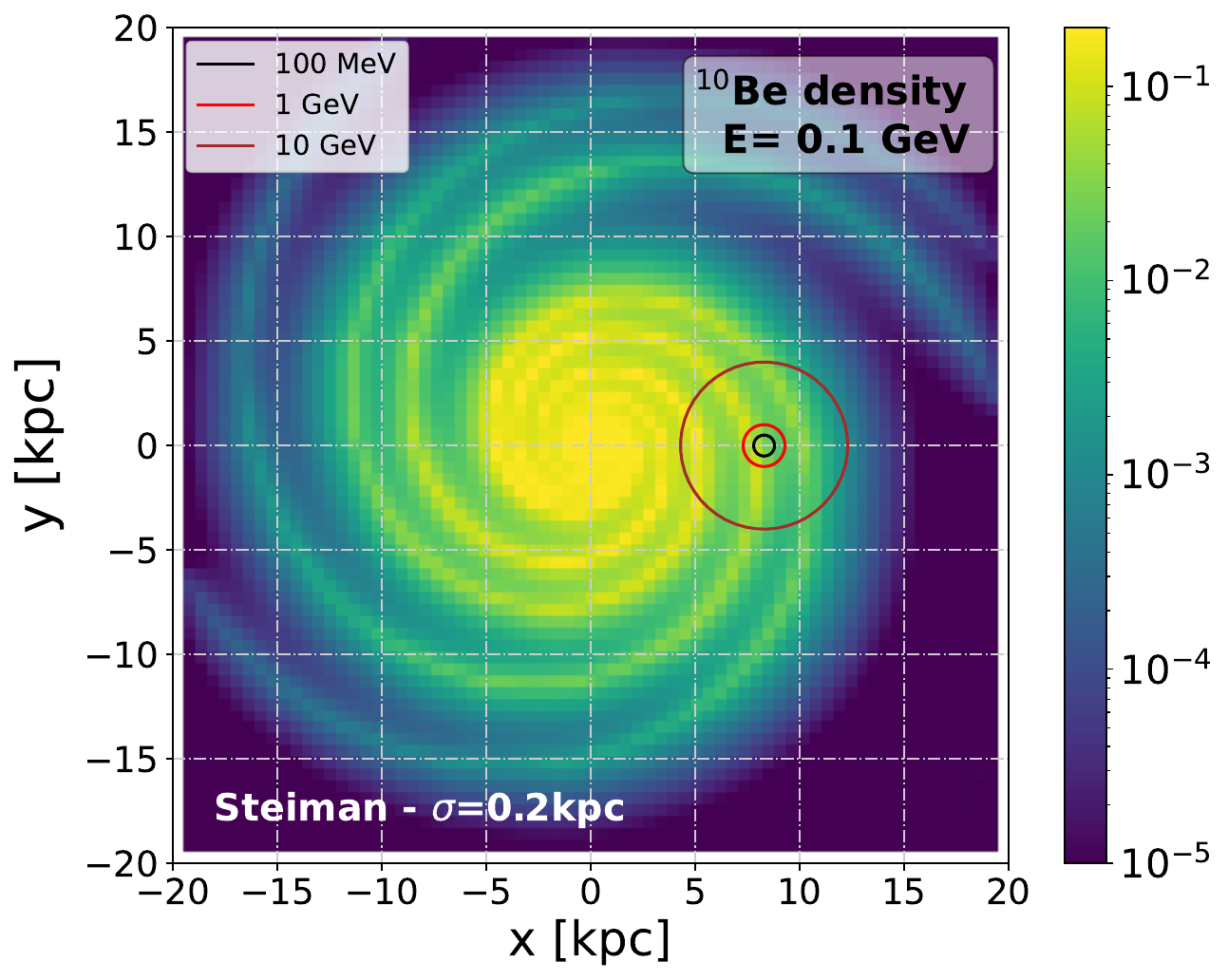} 
\hspace{-0.15cm}
\includegraphics[width=0.496\textwidth] {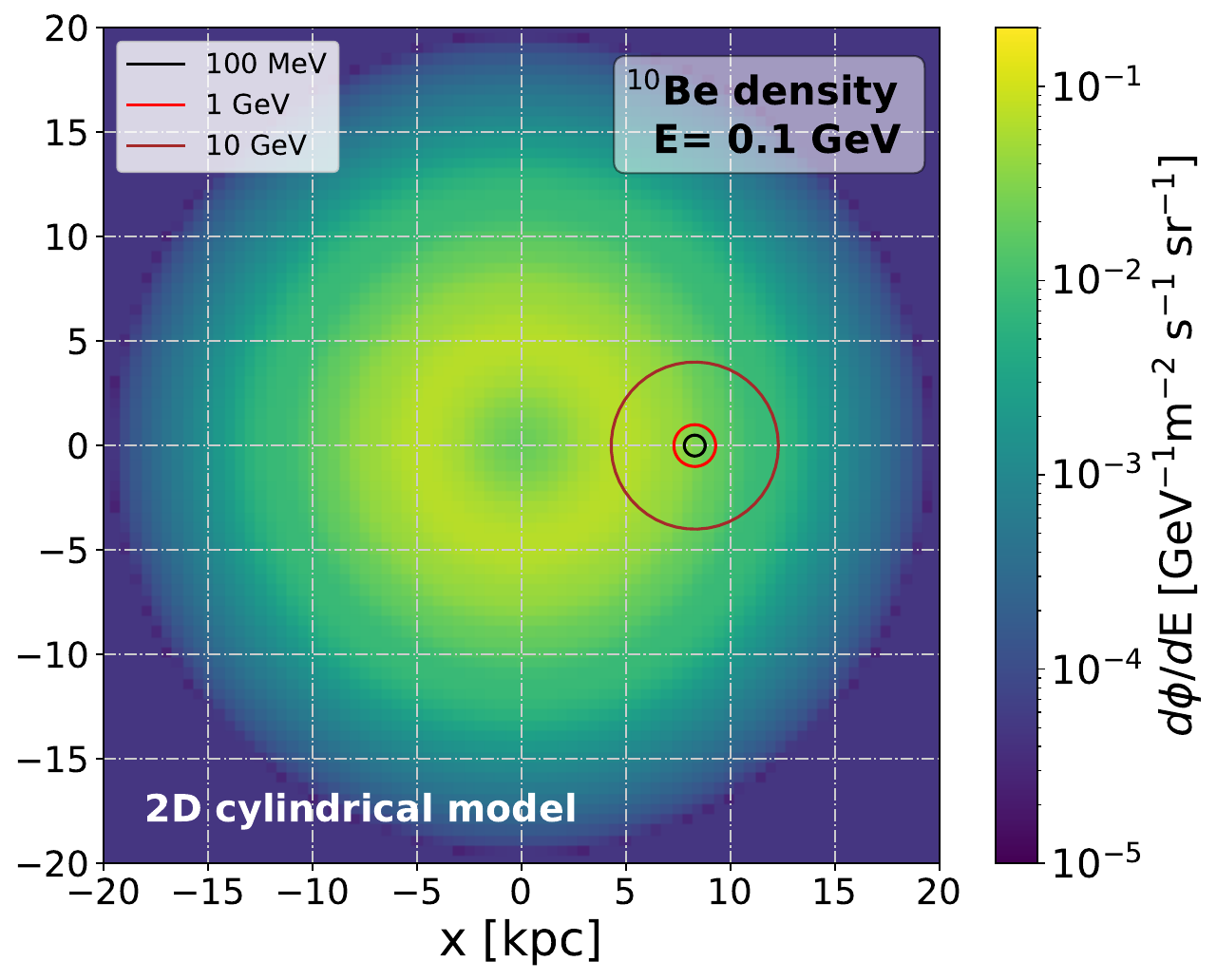} 

\caption{Predicted steady-state density distribution of $^{10}$Be at $0.1$~GeV {\bf after propagation} for a model adopting a 3D spiral arm structure of Galactic gas (left) or an azimuthally symmetric 2D gas model (right). The spiral arm model is from Steiman-Cameron et al~\cite{Steiman-Cameron_2010} with an arm width ($\sigma$) of $0.2$~kpc, while the 2D model follows work from the {\tt GALPROP} team~\cite{Strong_ICRC, Strong_1998}. Circles around the Sun represent the average distance travelled by a $0.1$~GeV, $1$~GeV, or $10$~GeV $^{10}$Be particle before decaying. 
}
\label{fig:Be10Maps}
\end{figure}

In this section, we demonstrate that moving from a 2D to 3D gas model systematically alters our predictions for the $^{10}$Be/$^{9}$Be ratio when all other diffusion parameters are fixed. In Fig.~\ref{fig:Be10Maps}, we illustrate the predicted $^{10}$Be flux in our 3D (left) and 2D (right) models. Even after CR propagation is taken into account, there is significant spatial structure in the $^{10}$Be flux in our 3D models, which is not accurately captured within any 2D model. The implication of this result, is that models of the $^{10}$Be flux sensitively depend on the accurate modeling of the closest spiral arms structures to Earth.

The limited propagation range of low-energy $^{10}$Be (at energies of $100$~MeV, $1$~GeV and $10$~GeV) is pictorally represented by circles of different colours that are centered on the Earth position. As one can see, below a GeV, this nucleus can not travel much more than $\sim2$~kpc, and thus the $^{10}$Be acts as a direct probe of the local environment. For concreteness, we calculate the mean distance travelled by this isotope as $\langle r \rangle (E) = \sqrt{2 D(E) \tau_{dec}(E)}$, where $D$ is the diffusion coefficient characterizing their motion in the Galaxy  and $\tau_{dec}$ is the decay lifetime ($\tau (E) = \gamma (E) \cdot \tau^0$, being $\tau^0$ the isotope lifetime at rest and $\gamma (E)$ the Lorentz factor). 

Critically, we note that the inhomogeneity of the $^{10}$Be flux cannot be corrected for by instead studying the isotopic ratio $^{10}$Be/$^{9}$Be. This is in stark contrast with most other degeneracies in the diffusion model. While the ratio $^{10}$Be/$^9$Be would not be affected by a constant change in the galactic gas density, the $^9$Be flux depends on the average gas density throughout the Milky Way, while the radioactive decays of $^{10}$Be make it extremely sensitive to the local environment. 
As Fig.~\ref{fig:Be10Maps} shows, the  $^{10}$Be density is much higher in the spiral arms than the average density predicted by the homogeneous disk model. This makes the local $^{10}$Be flux highly sensitive to the location of the Earth with respect to the spiral arms. 

\begin{figure}[!t]
\centering
\includegraphics[width=0.495\textwidth, height=0.23\textheight] {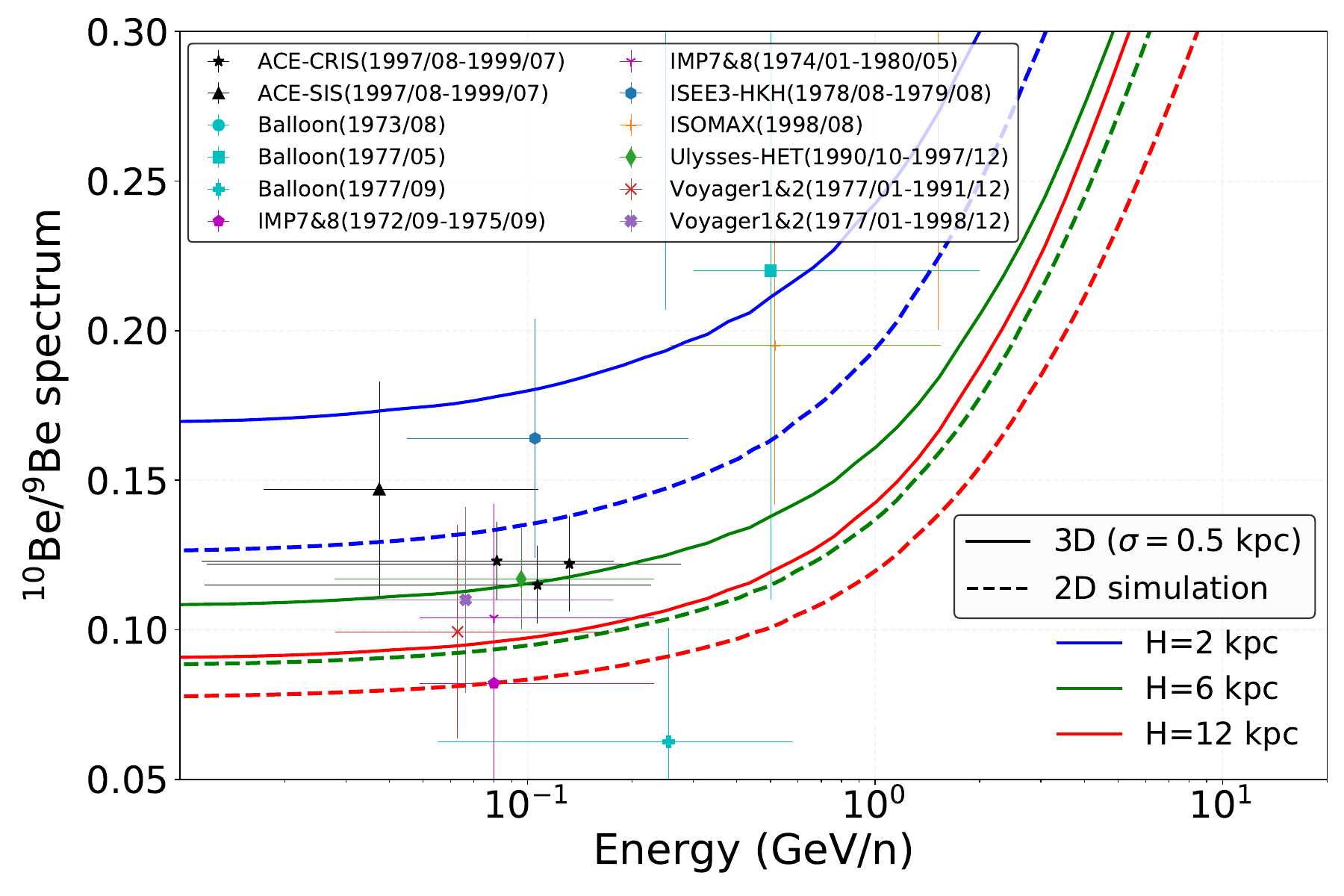} 
\includegraphics[width=0.495\textwidth,  height=0.22\textheight] {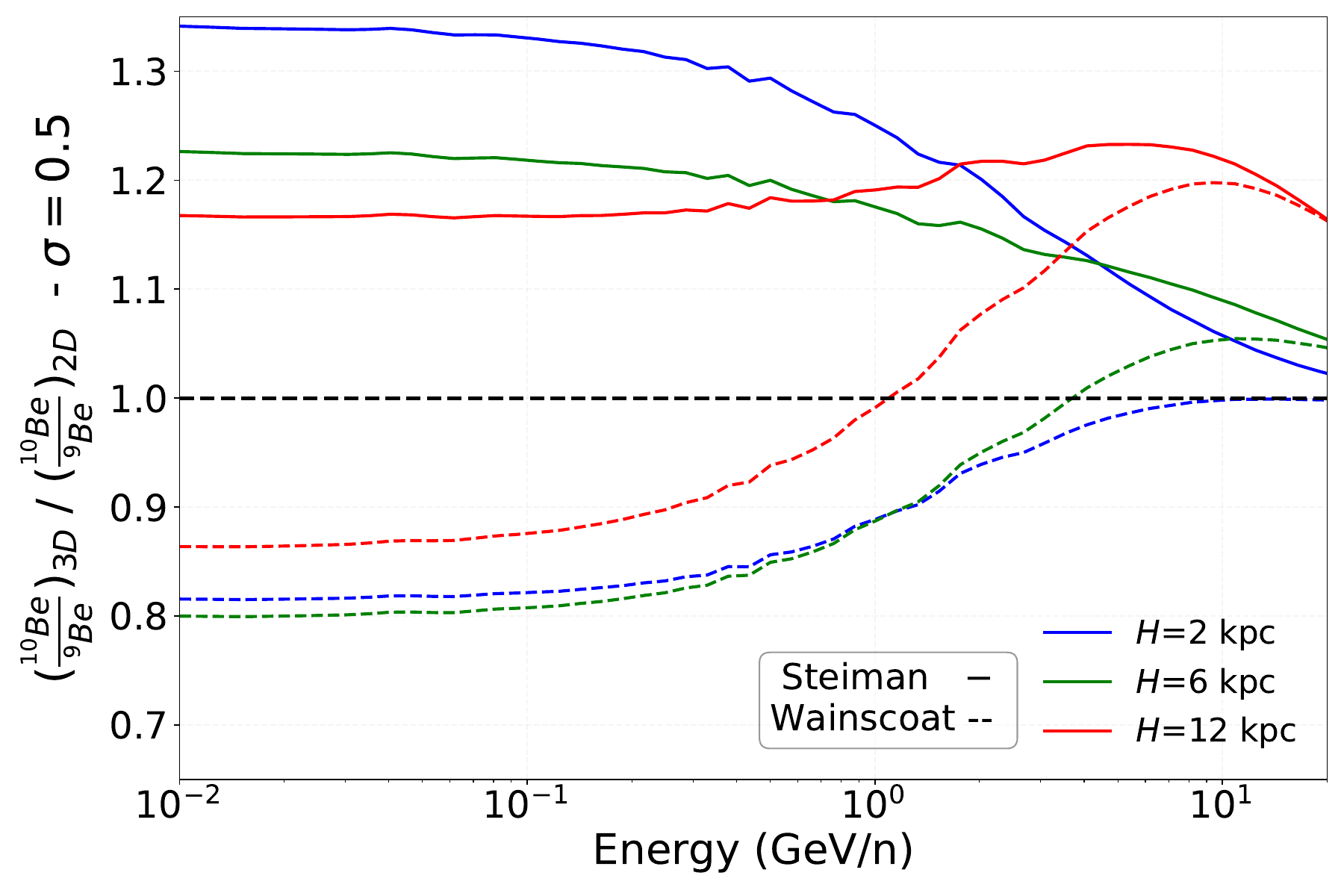} 
\caption{\textbf{Left panel:} The predicted $^{10}$Be/$^{9}$Be flux ratio for the Steiman-Cameron (3D) model of the Milky Way spiral arms (solid lines) and the {\tt GALPROP} 2D gas density model (dashed lines) for different halo heights.  \textbf{Right panel:} Relative difference in the predicted $^{10}$Be/$^{9}$Be ratio for the adoption of two 3D spiral arm gas density models (Wainscoat and Steiman) with respect to the ratio from the {\tt GALPROP} 2D density model. The model is fit to reproduce AMS-02 B/C data in each case.} 
\label{fig:Be10_2Dvs3D}
\end{figure}

To illustrate this point, Fig.~\ref{fig:Be10_2Dvs3D} (left) shows the $^{10}$Be/$^9$Be ratio when we adopt our benchmark 3D and 2D models for different values of the galactic halo height. The primary result of our work is that changing from a 2D to 3D gas model significantly changes the predicted $^{10}$Be/$^9$Be ratio in a way that is degenerate with changes in the halo height. Thus, studies that employ a 2D gas model to fit $^{10}$Be/$^9$Be will systematically prefer an incorrect halo height, compared to realistic 3D models. We note that this effect can be huge, a 3D gas model with a 6~kpc halo predicts almost the same $^{10}$Be/$^9$Be ratio as a 2D gas model with a 12~kpc halo.

We note that, for every halo height, we adjust the diffusion coefficient such that we reproduce the AMS-02 data for stable CR fluxes, as shown in Figs.~\ref{fig:Prims} and~\ref{fig:Secs} (App.~\ref{sec:AppendixC}) 
This allows us to regulate the grammage traversed by CRs, which is proportional to $H/D$ and fit well-measured quantities like B/C. However, since these are ratios of stable species, they track the average galactic grammage, while $^{10}$Be is especially sensitive to the very local ISM.

Fig.~\ref{fig:Be10_2Dvs3D} (right) displays the ratio between the predicted $^{10}$Be/$^{9}$Be spectrum for our 3D models compared to a 2D model with the same halo height. The ratio obtained using our benchmark spiral arms model (Steiman) is shown with solid lines, while the alternative Wainscoat distribution is shown with dashed lines. 
Importantly, we find that for the Steiman case, we expect a higher $^{10}$Be/$^{9}$Be ratio than in the 2D model because of the higher gas concentration near Earth, while the Wainscoat model predicts a lower $^{10}$Be/$^{9}$Be ratio than the 2D case, because the closest spiral arm lies farther from Earth. These results imply that accurate 3D spiral arms models are needed to accurately predict the $^{10}$Be/$^{9}$Be ratio. 
 
We note that at $100$~MeV, where most measurements lie, the difference in the $^{10}$Be/$^{9}$Be ratio for the 3D versus 2D model is between $\sim$17--33$\%$, for the Steiman model, and between $\sim - 12\%$ and $\sim -20\%$, for the Wainscoat model (see Table~\ref{tab:HComps}, in App.~\ref{sec:AppendixB}, for differences at various energies and halo heights). These quantities slightly differ for different values of the spiral arm width, $\sigma$, being slightly larger for a smaller $\sigma$ (see the discussion in App.~\ref{sec:AppendixB}), but the overall conclusion is the same in each case. We observe that, around $\sim10$~GeV, the differences on the predicted $^{10}$Be/$^{9}$Be ratio converge for all models. This is due to the fact that $^{10}$Be propagates significantly longer distances before decaying. 

Other unstable nuclei have been discussed as CR clocks, and will suffer from similar uncertainties as $^{10}$Be. We have implemented the propagation of $^{26}$Al in DRAGON2, having taken into account the probability of capturing an electron from the ISM and its subsequent decay. We find that the local $^{26}$Al/$^{27}$Al ratio suffers from similar variations as the $^{10}$Be/$^{9}$Be ratios whe adopting different gas distributions, which we show in Fig.~\ref{fig:Al_3D_2D} of App.~\ref{sec:AppendixB}.


\subsection{Effect of the local bubble in our evaluations}
\label{sec:LocalBubble}

\begin{figure}[!t]
\centering
\includegraphics[width=0.494\textwidth] {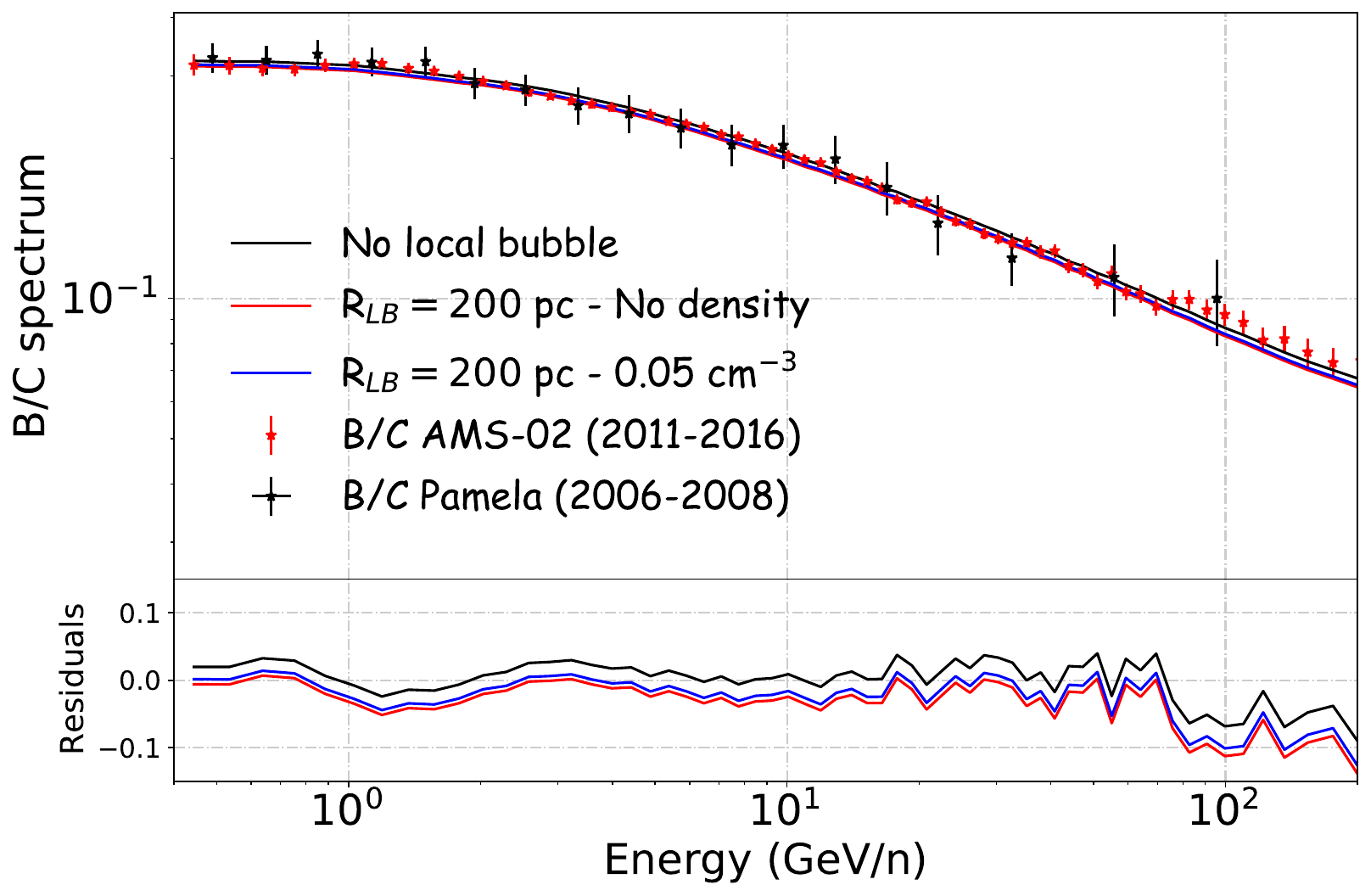} 
\includegraphics[width=0.494\textwidth] {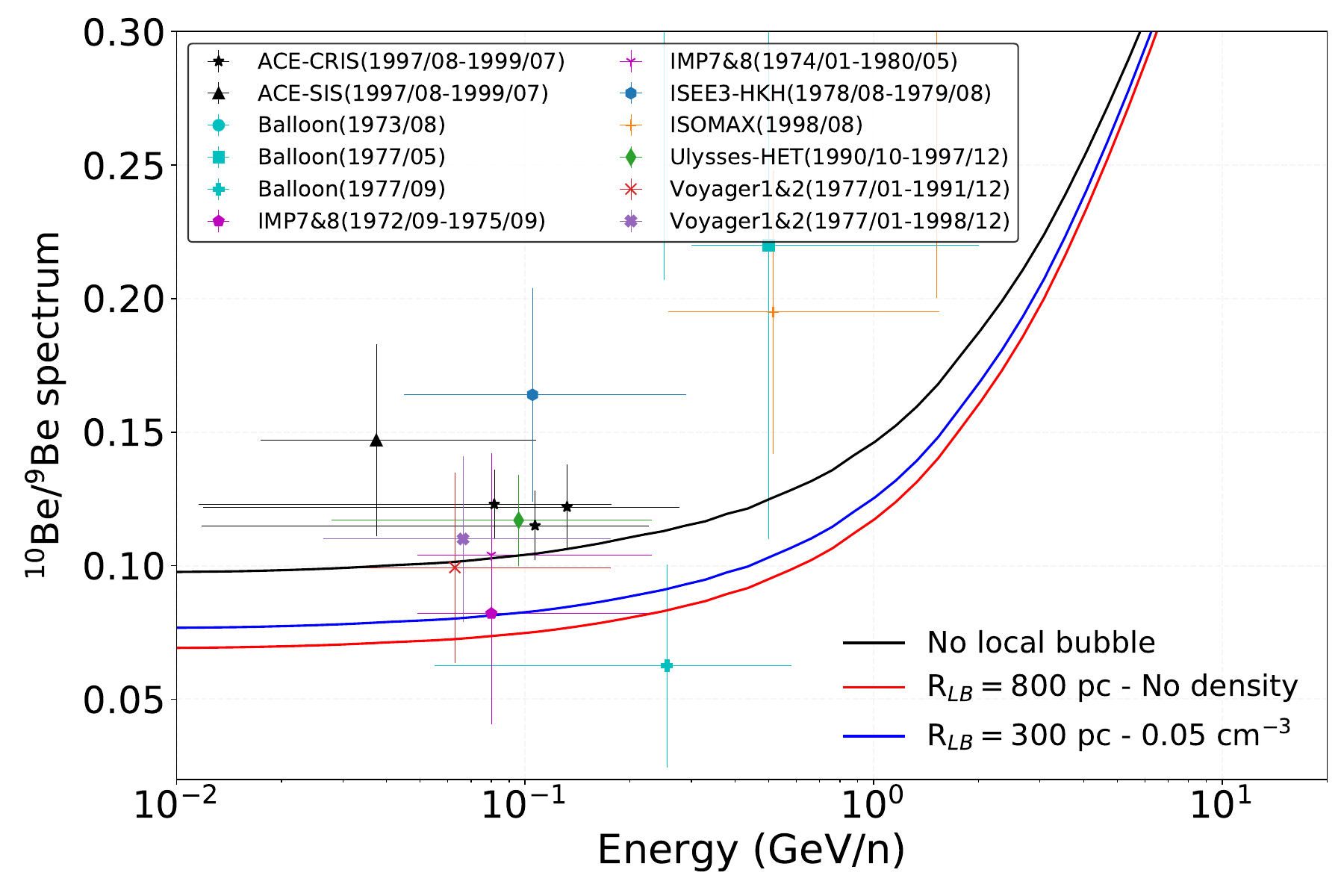} 
\caption{Effect of a local bubble in the predicted local B/C ratio (left panel) and $^{10}$Be/$^{9}$Be ratio (right panel). We compare the cases in which there is no local bubble, the case of a local bubble with $1/20$ times the expected local density and the case of no density inside the bubble. }
\label{fig:Bubble_Comp}
\end{figure}

There is robust evidence of an underdense ($\sim$20$\times$ less dense that the average Milky Way gas density) ``bubble" that contains the Solar System and is thought to be the product of a series of supernovae occurring between $10$ and $20$~Gyr ago. The impact of the local bubble on the fluxes of unstable CRs has been studied in the past~\cite{2015ICRC...34..543G, Donato_2002}, however it has typically been found to only negligibly affect most CR observables. This, combined with the uncertain extent of the ``bubble", have caused many studies to ignore bubble contributions in their fits. However, our results indicate that the local bubble may uniquely affect the predicted $^{10}$Be flux at Earth, and therefore our estimate of the Galactic halo height. 

Here, we implement a spherical bubble, centered at Earth, with a radius of $200$~pc and a gas density that is underdense by a factor of $\sim20$ (i.e. the gas density within $200$~pc from Earth is $20$~times lower than in the local interstellar medium around it), roughly corresponding to a density of $0.05$~cm$^{-3}$. In this study, we assume that diffusion throughout the bubble is otherwise identical to the surrounding ISM, and leave a more detailed study about the impact of the local bubble, including possible variations of the diffusion coefficient~\cite{Gebauer_Bubble}, to future work. We also note that our analysis could used a spatial resolution just a bit lower than $100$~pc, which must be improved to more accurately study diffusion in the local bubble.

In Fig.~\ref{fig:Bubble_Comp}, we compare the predicted local B/C (left panel) and $^{10}$Be/$^{9}$Be (right panel) ratios for the benchmark case of a Steiman spiral arm distribution with $\sigma= 0.5$~kpc and $H=8$~kpc (because it provides a good fit to the $^{10}$Be/$^{9}$Be and $^{10}$Be/Be data), without the local bubble (black lines), with a local bubble of density $\sim0.05$~cm$^{-3}$ and with a local bubble that has no gas (red lines), which represents an exaggerated case. While we see, from the left panel, the impact of including a bubble is negligible for the B/C ratio (the relative difference in the case of no bubble even with the case of no density inside the bubble is roughly $1\%$), its impact is more important for the $^{10}$Be/$^{9}$Be ratio, where the ratio for the no bubble case is around $20\%$ higher than in the case of the local bubble with $\sim0.05$~cm$^{-3}$ density and more than $30\%$ than in the case that there is no gas within the local bubble.

Given that there are no other important CR features that can allow us to discern between the model with and without bubble, this results in another source of degeneracy when determining the value of the halo height from $^{10}$Be/$^{9}$Be measurements.

\section{Discussion on other sources of uncertainties}
\label{sec:OtherUncerts}

Previous works have argued that the uncertainties in cross sections measurements are much larger than current errors in CR flux measurements, leading to significant systematic uncertainties in the determination of propagation parameters such the halo height~\cite{GenoliniRanking, Derome_2019, Weinrich:2020cmw, Tomassetti:2015nha, Maurin_Be10, Luque:2021joz, Luque_MCMC, delaTorreLuque:2022vhm, Korsmeier:2021brc, zhao2024interpretation}. These uncertainties have been typically quantified to be around $20-30\%$, which is enough shift best-fit halo heights over a range between $\sim 3$~kpc and $>10$~kpc. We remark that this level of uncertainty is similar to the variations that our gas distribution models can produce in our evaluations unstable isotopic ratios.

Different approaches have been used to account for cross sections uncertainties. Refs.~\cite{Luque:2021joz, Luque_MCMC} include scaling factors that are penalized proportionally to the mean error in cross section measurements, Refs.~\cite{Korsmeier:2021bkw, Korsmeier:2021brc} include free scaling factors above $5$~GeV/n and a parameter to modify the cross section energy dependence at lower energies, Refs.~\cite{Weinrich_halo, Weinrich:2020cmw, Derome_2019} use covariance matrices in their analysis, while Ref.~\cite{zhao2024interpretation} used different scaling factors at low and high energies to account for these uncertainties. Importantly, we note that these analyses mainly modify a given cross section parameterization to account for uncertainty in the measurements, but the adoption of any parametrization already introduces a systematic bias.

\begin{figure}[!t]
\centering
\includegraphics[width=0.495\textwidth] {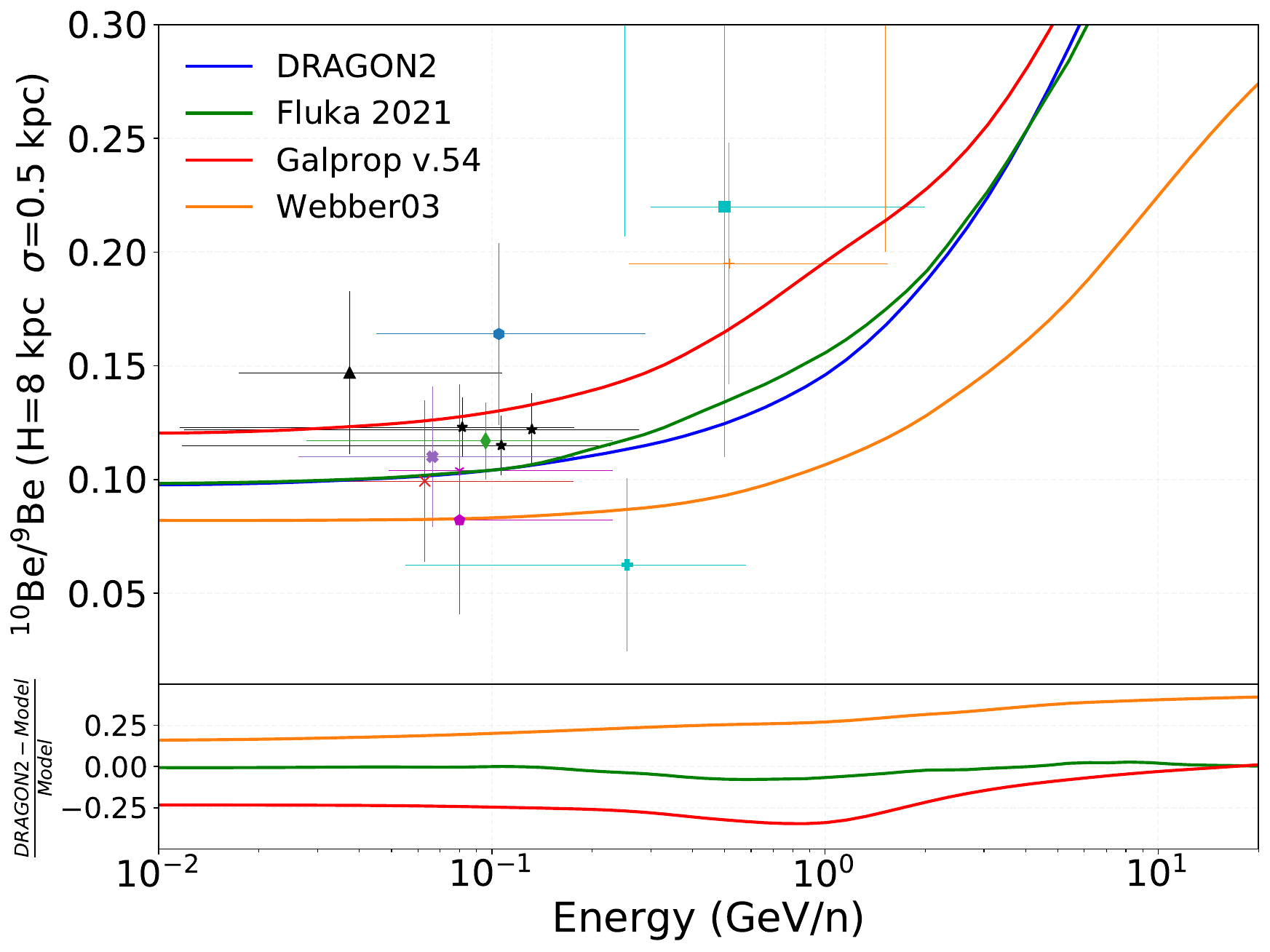} 
\includegraphics[width=0.495\textwidth] {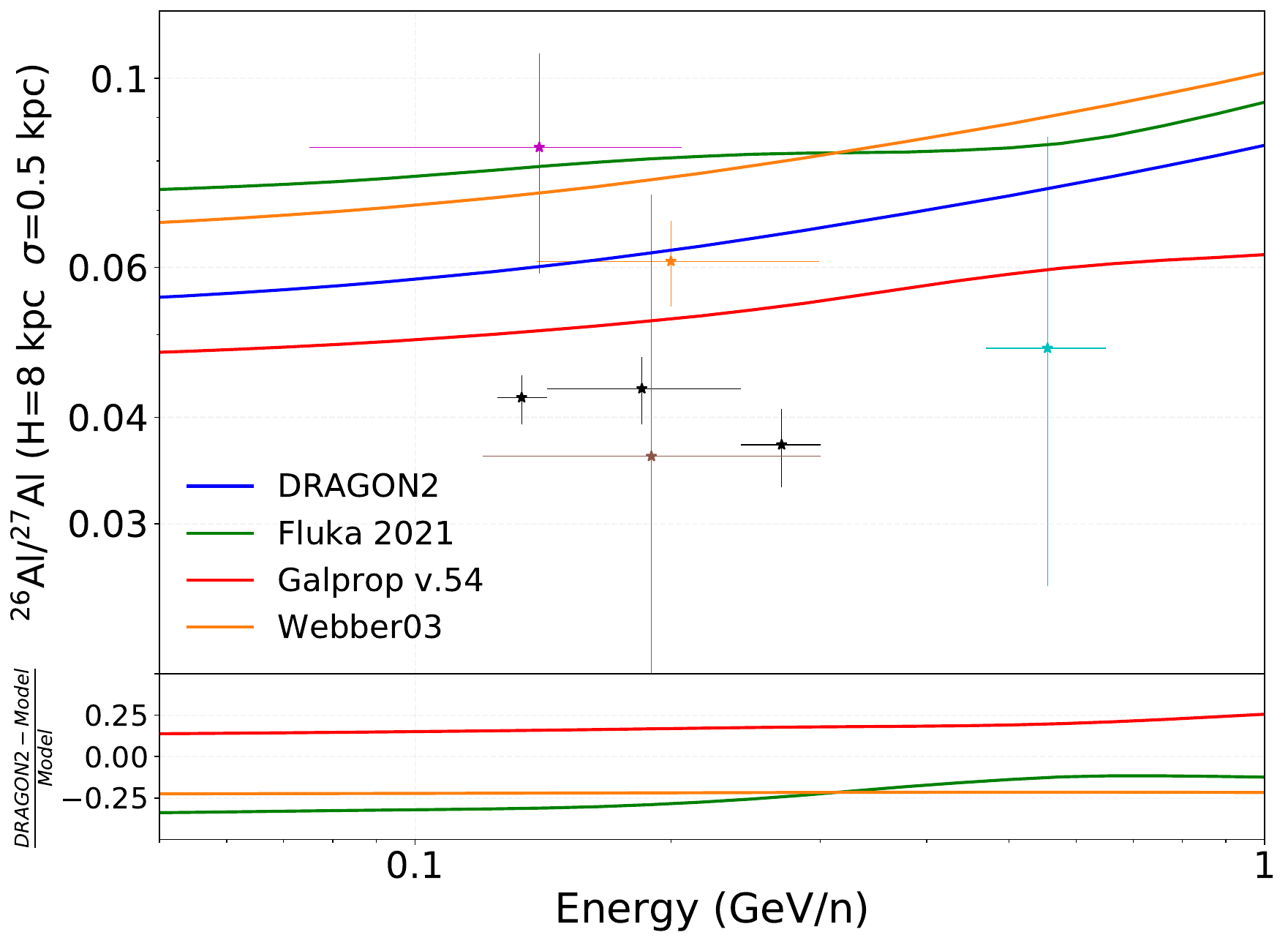}
\caption{Comparison of the predicted $^{10}$Be/$^{9}$Be (left panel) and $^{26}$Al/$^{27}$Al (right panel) for different popular cross sections parameterizations. AMS-02 data is reproduced in every case.}
\label{fig:XS}
\end{figure}

In the left panel of Fig.~\ref{fig:XS}, we show the predicted $^{10}$Be/$^{9}$Be ratio for different popular cross sections parameterizations (namely GALPROP~\cite{strong2007cosmic, moskalenko2001new, moskalenko2005propagation}, DRAGON2~\cite{Evoli:2019wwu, DRAGON2-2, Luque:2021joz}, FLUKA~\cite{delaTorreLuque:2022vhm, Mazziotta:2020uey, DelaTorreLuque:2023zyd, DelaTorreLuque:2022czl} and Webber (WNEW03)~\cite{webber1990formula, webber2003updated}), for a gas distribution following the Steiman spiral arm model and $H=8$~kpc. The relative differences in each model with respect to the DRAGON2 cross sections are also shown below. In all cases, we adopt propagation parameters that reproduce the local AMS-02 B/C spectrum. In the right panel of this figure, we show the same comparison but for the $^{26}$Al/$^{27}$Al ratio.
In both panels, we see that the predicted differences are within $\sim30\%$, as typically reported. Recent campaigns are devoted to improve the quality of cross section data for astrophysical applications, from the production of antiprotons and antinuclei, to heavier nuclei~\cite{gondolo2013wanted, GenoliniRanking, Doetinchem_2020}, therefore, we expect that this source of uncertainty can soon be reduced. However, we want to emphasize here that while the impact of cross sections uncertainties on $^{10}$Be predictions have received significant attention in recent work, the impact of not adopting a realistic gas density distribution can be similar or even greater, indicating that its careful treatment requires a similar effort.

In addition, uncertainties from solar modulation and the determination of the diffusion parameters have been also discussed and analyzed recently. In Ref.~\cite{Weinrich_halo}, the authors show that uncertainties on the $^{10}$Be ratios from solar modulation are of the order of a few percent. However, this analysis only accounted for uncertainties in the determination of the Fisk potential under the Force-Field approximation~\cite{ForceField}. Other approaches to study solar modulation could introduce additional uncertainties in the energy range where current data are available. Energy losses or uncertainties in the estimation of the diffusion parameters are similarly expected to be small for heavy nuclei at low energies.

\section{Conclusions}
\label{sec:conc}

In this paper, we demonstrate that accurately modeling the local 3D distribution of galactic gas produces a significant and systematic shift in the $^{10}$Be/$^{9}$Be and $^{26}$Al/$^{27}$Al isotopic ratios, which have historically been discussed as the optimal radioactive secondary ratios to constrain galactic diffusion. In particular, utilizing 3D gas maps has a significant effect on the evaluation of the Galactic halo height, which is a crucial parameter not only for CR physics, but in particular for indirect DM searches.



The systematic shift depends on the Spiral arms model. Here we have used, as a benchmark, the Steiman-Cameron 4-arm model, for which variations, with respect to the homogeneous model of up to more than $30\%$ (higher ratios) have been found at $100$~MeV. Using a different spiral arm model, the Wainscoat model, we have obtained similar variations of up to -$20\%$, where the 3D model predicts lower ratios than in the 2D case.
Unfortunately, we find that the different gas distributions produce no other significant features, therefore making different choices of gas distribution highly degenerated with the halo height. Additionally, we show that similar but slightly smaller systematic shifts are produced in models that include the under-dense local bubble around the Solar System, which similarly does not appreciably impact other CR observables.

These variations in the predicted $^{10}$Be/$^{9}$Be ratio are similar to those produced from the choice of different cross sections parameterizations, leading to another important uncertainty in our models of the halo height. In fact, our analysis shows that well-motivated modles of the gas distribution could be in good agreement with the current data for any value of $H$ between $\sim3$ and $\sim12$~kpc.
We note that this study is not intended to be a precise evaluation of the uncertainties related to the gas distributions, but to illustrate that this is another important source of systematic uncertainty in our evaluations of the galactic halo height, with an effect that is at least as significant as those from cross-section measurements. 

In sum, we show that the importance of the gas density distribution in the evaluation of $^{10}$Be ratios mean that a robust determination of the halo height requires a detailed modelling of the spiral arm structure and the local bubble in CR propagation codes. Having data at higher energies could help reduce the uncertainties, given that the impact of the spiral arm structure of the gas density becomes less important at higher energies. Upcoming measurements of these ratios from AMS-02 and future measurements from HELIX may be critical in providing mores stringent constraints on galactic isotopic ratios.

The main input and output DRAGON files used in this work are available at \url{https://github.com/tospines/Codes-Models-ExtraPlots/tree/main/10Be_GasDistribution}.

\acknowledgments
P.D.L. is currently supported by the Juan de la Cierva JDC2022-048916-I grant, funded by MCIU/AEI/10.13039/501100011033 European Union "NextGenerationEU"/PRTR. The work of P.D.L. is also supported by the grants PID2021-125331NB-I00 and CEX2020-001007-S, both funded by MCIN/AEI/10.13039/501100011033 and by ``ERDF A way of making Europe''. P.D.L. also acknowledges the MultiDark Network, ref. RED2022-134411-T.  T.L. acknowledges support by the Swedish Research Council under contract 2022-04283.This project used computing resources from the National Academic Infrastructure for Supercomputing in Sweden (NAISS) under project NAISS 2023/3-21. During the beginning of this paper, P.D.L. was supported by the Swedish National Space Agency under contract 117/19. We acknowledge the {\tt FLUKA} collaboration for providing and supporting the code.

\bibliographystyle{apsrev4-1}
\bibliography{biblio}

\newpage
\appendix

\section{Galactic $^{10}$Be distribution map}
\label{sec:AppendixA}
\begin{figure}[!h]
\centering
\includegraphics[width=0.495\textwidth] {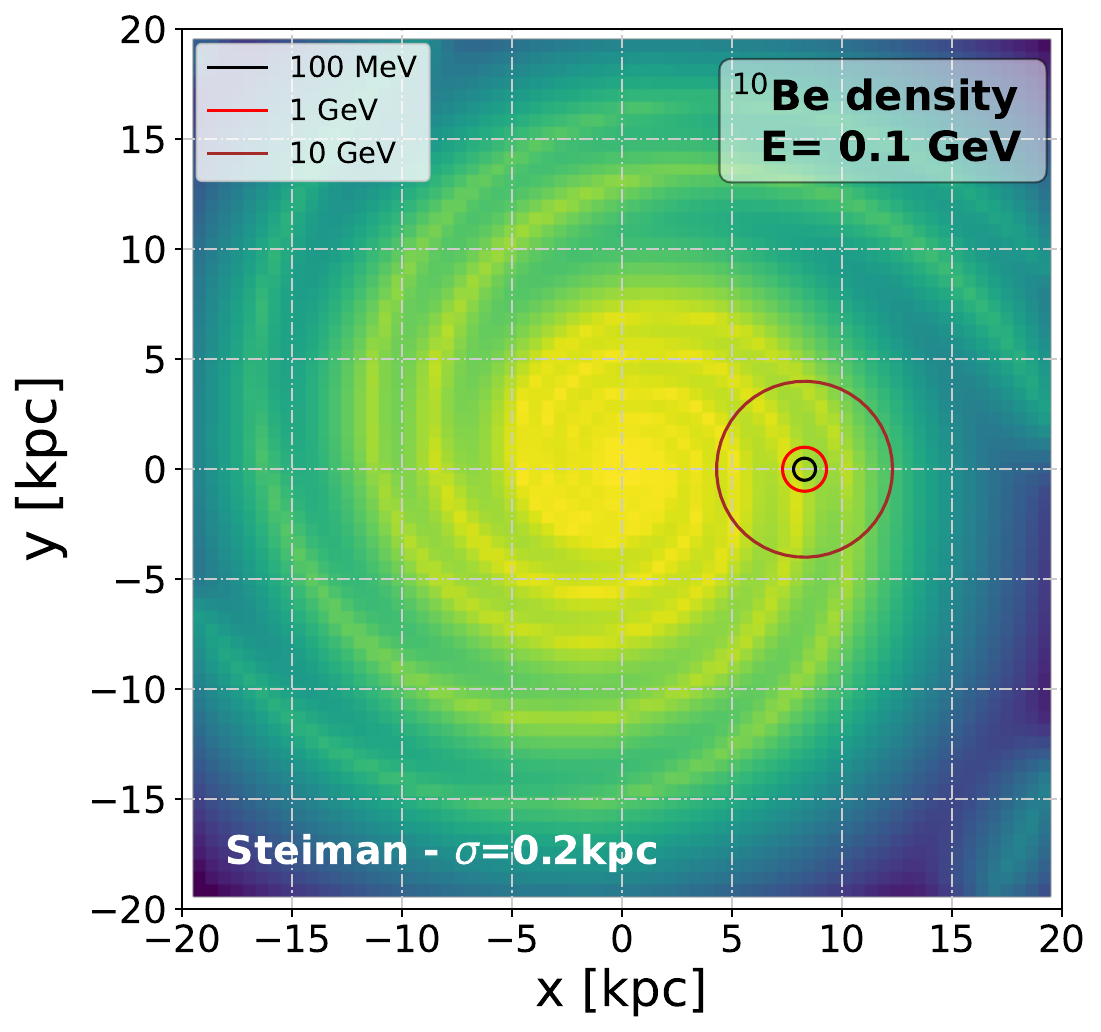}  \hspace{-0.2cm}
\includegraphics[width=0.495\textwidth] {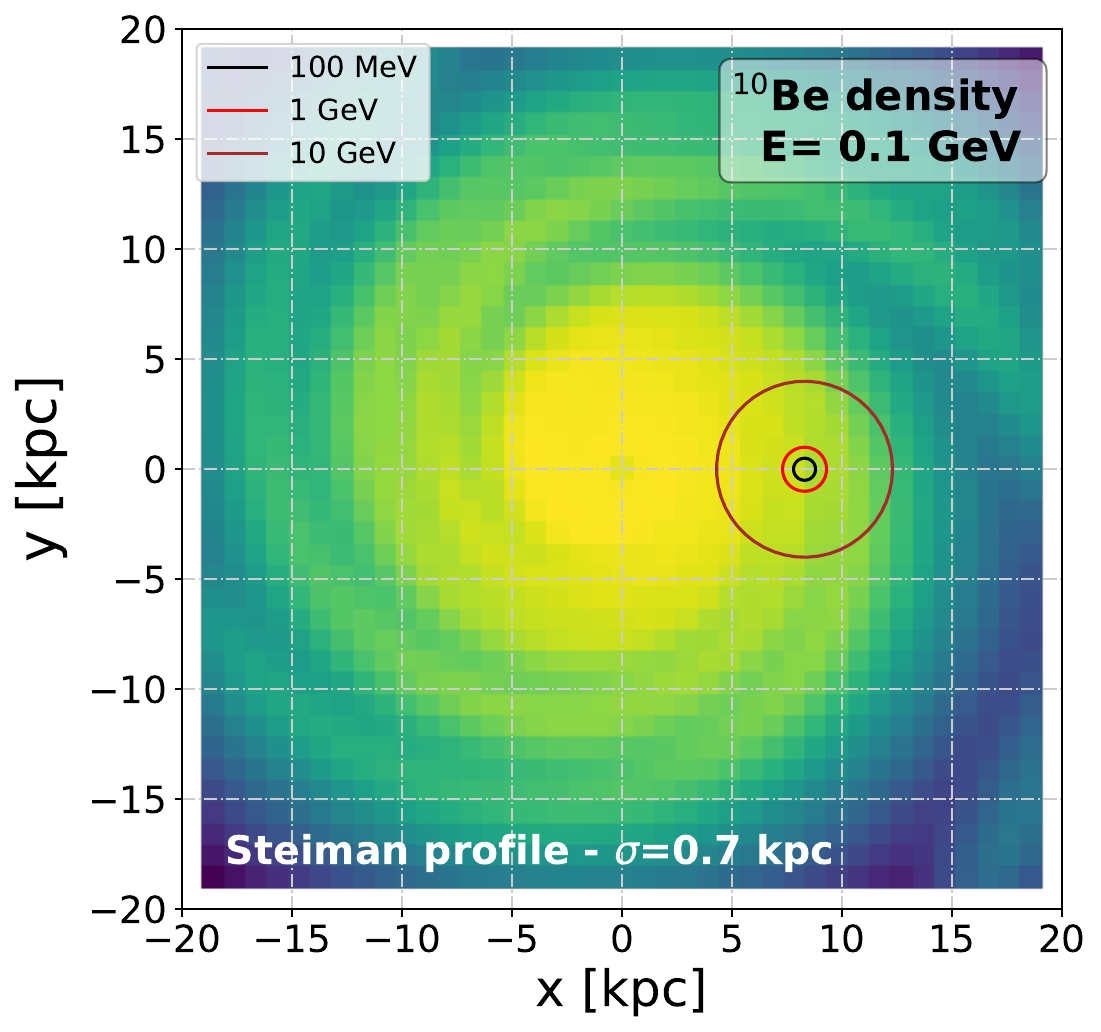} 
\vskip 0.2in
\includegraphics[width=0.495\textwidth] {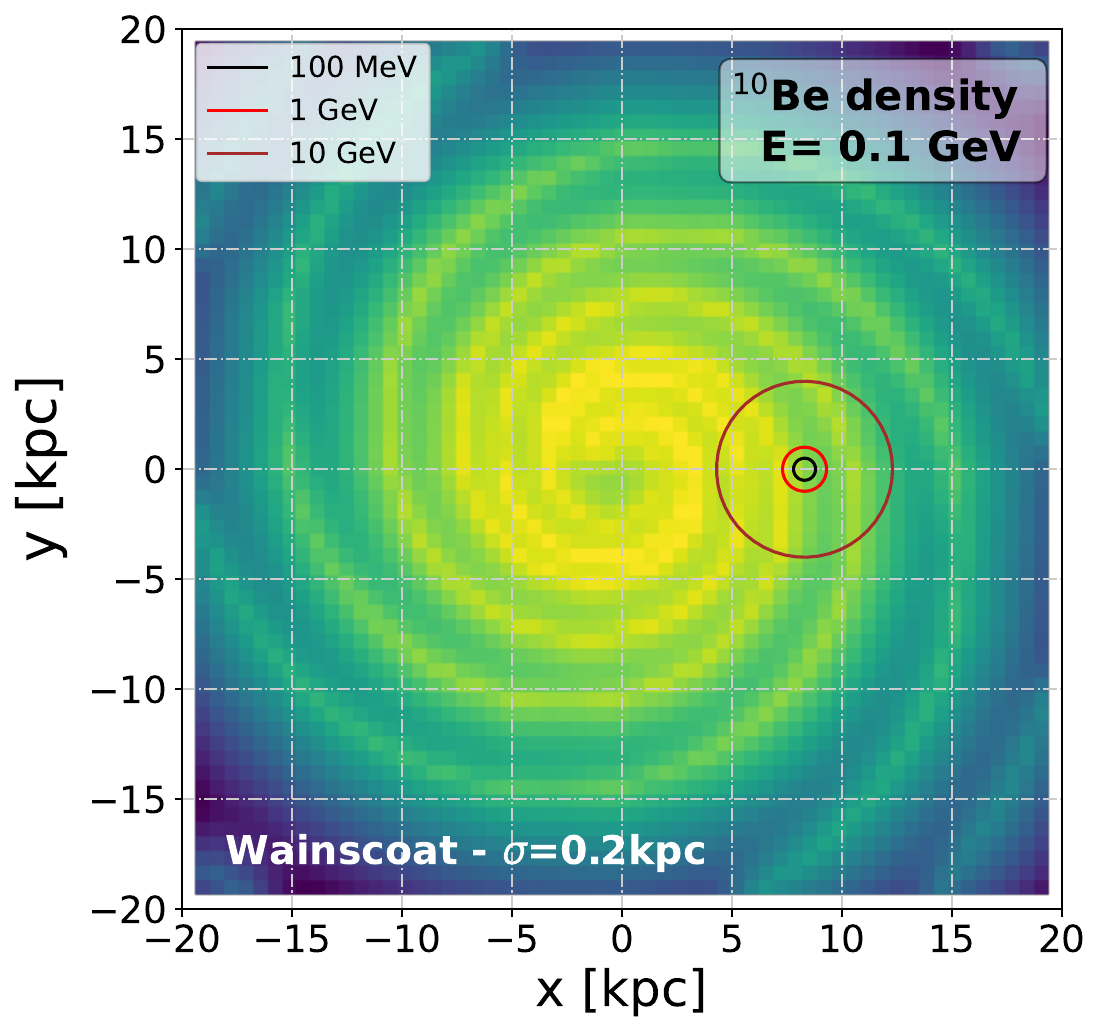} \hspace{-0.2cm}
\includegraphics[width=0.495
\textwidth] {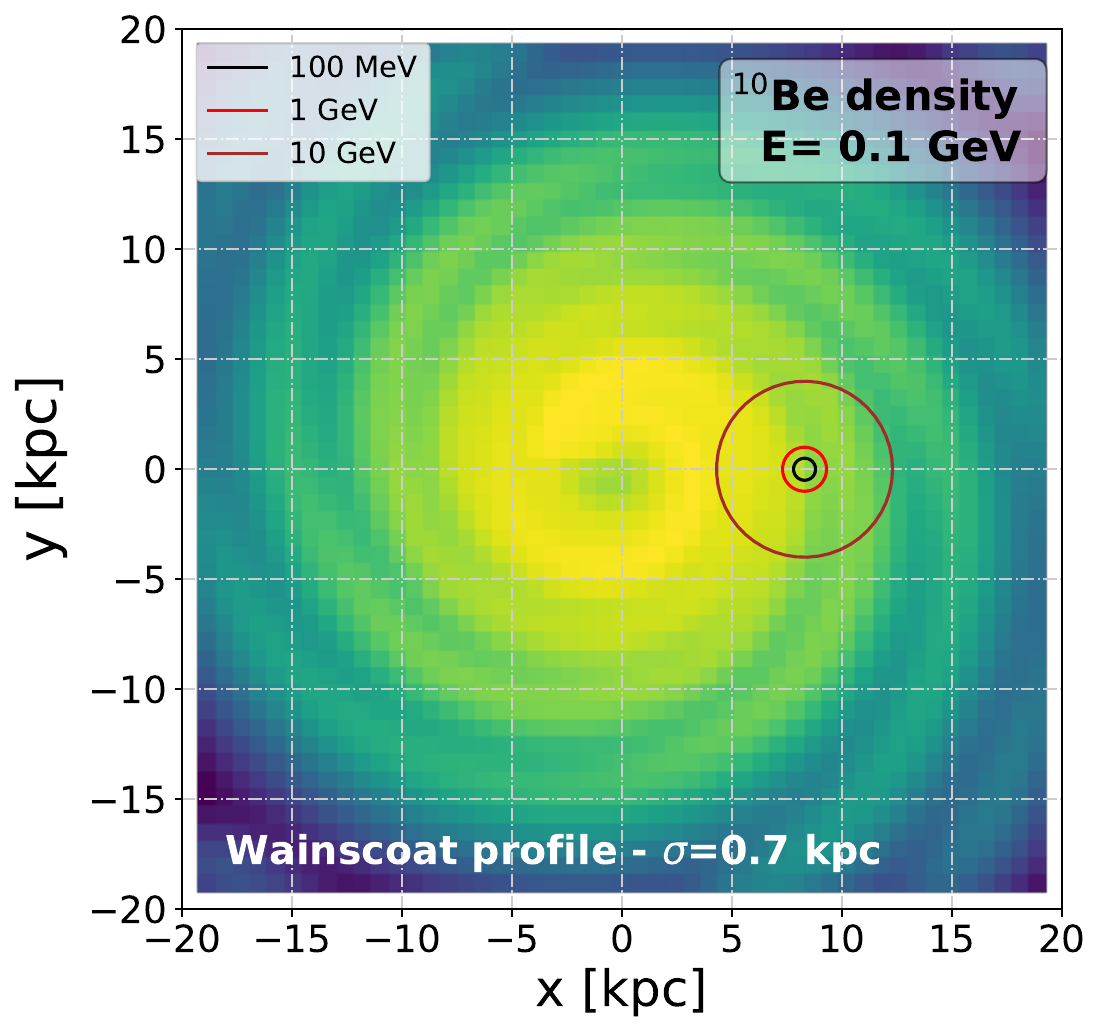} 
\caption{Comparison of the steady-state $^{10}$Be density for different configurations of the spiral arm models used. Top panels show the density predicted from the Steiman-Cameron 4-arm model and the bottom panels the density obtained adopting the Wainscoat model. The results are shown for a spiral arm width, $\sigma$, of $0.2$~kpc (left panels) and $0.7$~kpc (right panels).}
\label{fig:Be10Maps_App}
\end{figure}

In Fig.~\ref{fig:Be10Maps_App} we display the predicted maps of the steady-state $^{10}$Be density for different configurations of the spiral arm models adopted in this work. These maps show interesting features that arise from differences in the gas density. In particular, in the top panels we show the density of $^{10}$Be evaluated from the Steiman-Cameron 4-arm model and, in the bottom panels, the density evaluated using the Wainscoat model. It is interesting to note the difference in the relative position of the Solar system with respect to the spiral arms. In the Wainscoat profile, the solar System lies outside the closest spiral arm, even in the case of arm width of more than $0.5$~kpc, while in the Steiman-Cameron model the Solar System is very close or inside the closest arm.

\section{Ratios of other species and additional details}
\label{sec:AppendixB}

This appendix discusses the effect of other ingredients involved in the modelling of the Galactic gas distribution on the predictions for unstable CR fluxes, and to provide additional figure showing the predicted ratios of $^{10}$Be/$^{9}$Be, $^{26}$Al/$^{27}$Al and the $^{10}$Be/Be and $^{10}$Be/$^{10}$B ratios.

\begin{figure}[!t]
\centering
\includegraphics[width=0.496\textwidth, height=0.25\textheight] {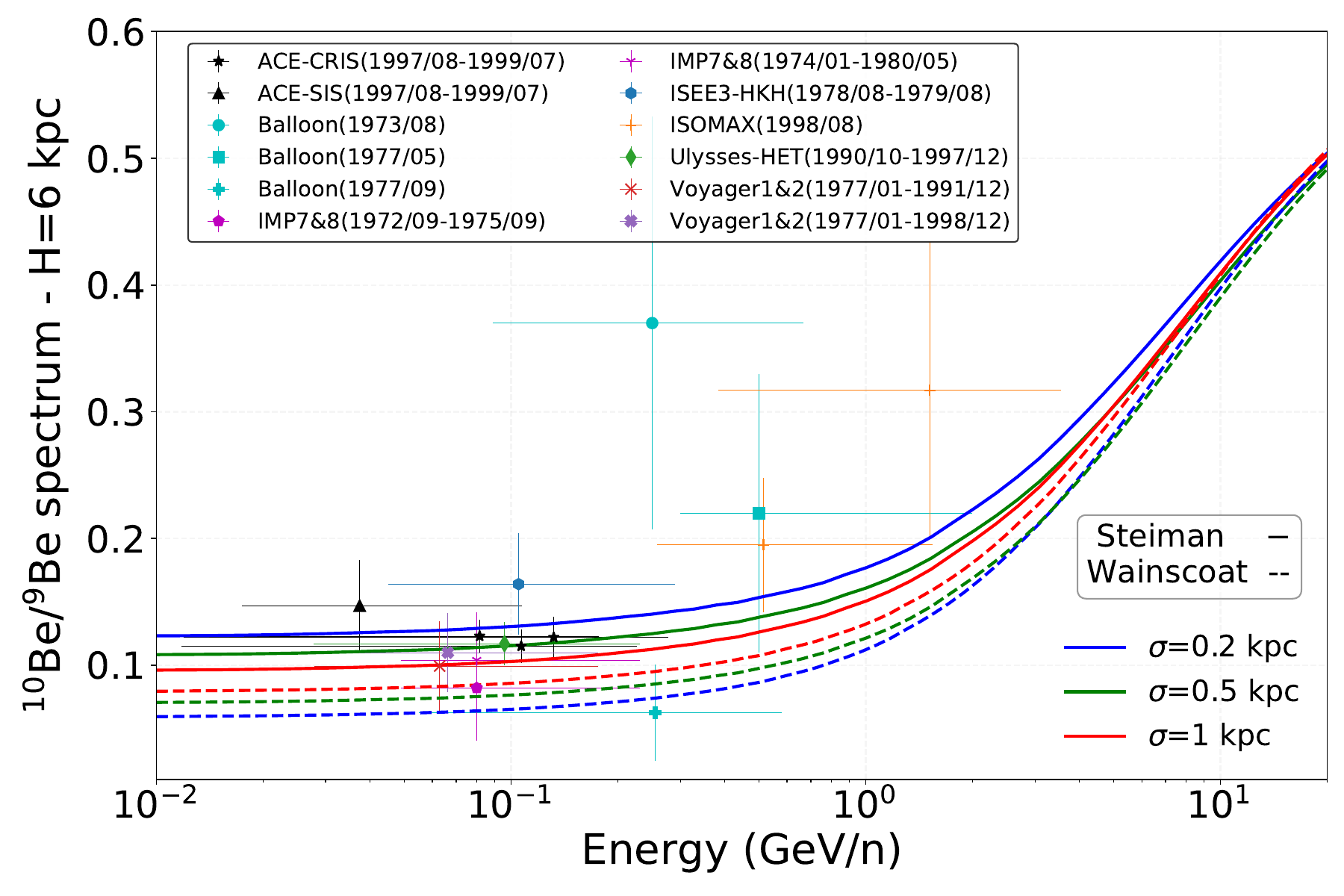} 
\includegraphics[width=0.496\textwidth, height=0.25\textheight] {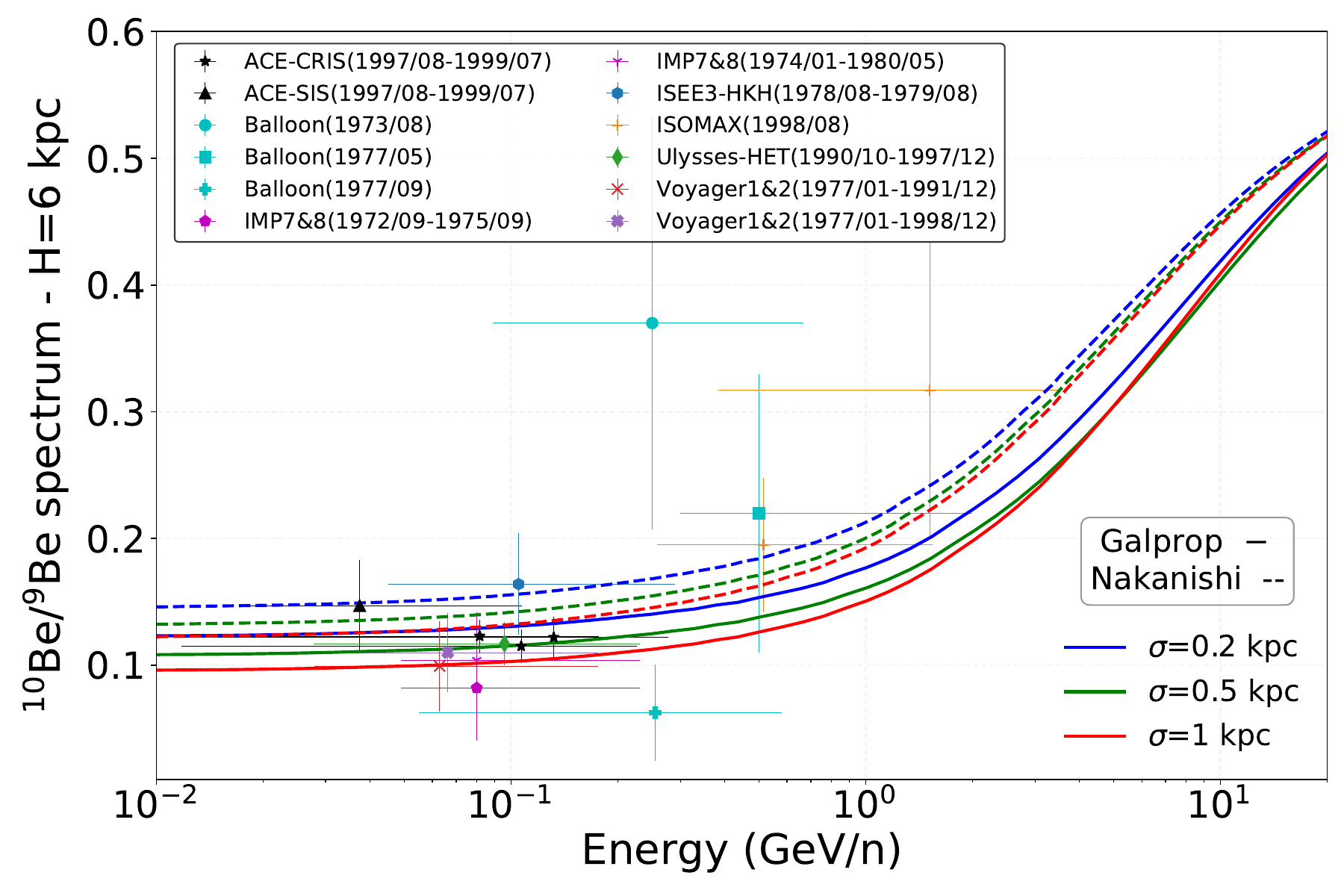} 
\vspace{-0.3cm}
\caption{Effect of variations in our benchmark spiral arm model on the local $^{10}$Be/$^{9}$Be ratio, for a halo height of $6$~kpc. \textbf{Left panel:} Effect of changing the width of the spiral arms ($\sigma$) in the predicted ratio for the Steiman-Cameron and the Wainscoat spiral arm models. \textbf{Right panel:} Effect of changing the radial dependence of the gas density, using the Galprop 2D model and the Nakanishi model, for different values of the spiral arm width.}
\vspace{-0.2cm}
\label{fig:Be10_Sigmas}
\end{figure}

In the main text we have compared the results for benchmark gas density models. These assume a given radial dependence and, in the 3D case, assume a fixed arm width of $\sigma= 0.5$~kpc. However, following the reasoning above, the spiral arm width will also determine the concentration of gas in the arms, which then affects the ratios of unstable isotopes at Earth. In the left panel of Fig.~\ref{fig:Be10_Sigmas}, we compare the predicted $^{10}$Be/$^{9}$Be ratio for different arm widths from $\sigma= 0.2$~kpc to $\sigma= 1$~kpc, assuming both our benchmark spiral arm model (Steiman model) and the Wainscoat model, for a fixed halo height of $H=6$~kpc (similar variations are found for other values of H). For completeness, we provide the relative differences between the two models, for different $\sigma$ values and energies, in Table~\ref{tab:SigmaComps}, in App.~\ref{sec:AppendixB}. As expected, the predicted ratio becomes closer to the 2D case as we use a larger arm width. 

In both 3D models, we note that the predicted $^{10}$Be/$^{9}$Be ratio varies by up to $\sim15\%$ when going from $\sigma= 0.2$~kpc to $\sigma= 1$~kpc, at $100$~MeV. Interestingly, we again observe opposite trends in how varying $\sigma$ affects the predicted ratio for the Steiman and Wainscoat models. Having the Solar System inside a spiral arm, as in the Steiman profile, one expects that more concentrated arms produce more $^{10}$Be that reaches the Solar System without decaying, for the same total grammage. However, in the Wainscoat profile, the Solar System lies outside the spiral arm, so that narrower spiral arms require longer travel distances from the center of the arm to the Earth, thus reducing the expected local $^{10}$Be/$^{9}$Be ratio.

Similarly, uncertainties in the radial gas density distribution also lead to uncertainties in the amount of $^{10}$Be produced around the Solar system. We illustrate this dependence in the right panel of Fig.~\ref{fig:Be10_Sigmas}, where we compare the $^{10}$Be/$^{9}$Be ratio obtained using the Galprop gas model and the Nakanishi gas distribution model~\cite{Nakanishi_2003}, for different values of $\sigma$ and for a halo height of $6$~kpc. The Nakanishi gas model features an azimuthally averaged surface density and scale height as function of galactocentric radius (see a full description in Sec.C1 of Ref.~\cite{DRAGON2-1}.
From this comparison, we see that one can expect differences of more than $10\%$ (see Table~\ref{tab:SigmaComps} for quantitative comparisons at different energies and different $\sigma$ values).

\begin{figure}[!t]
\centering
\includegraphics[width=0.496\textwidth] {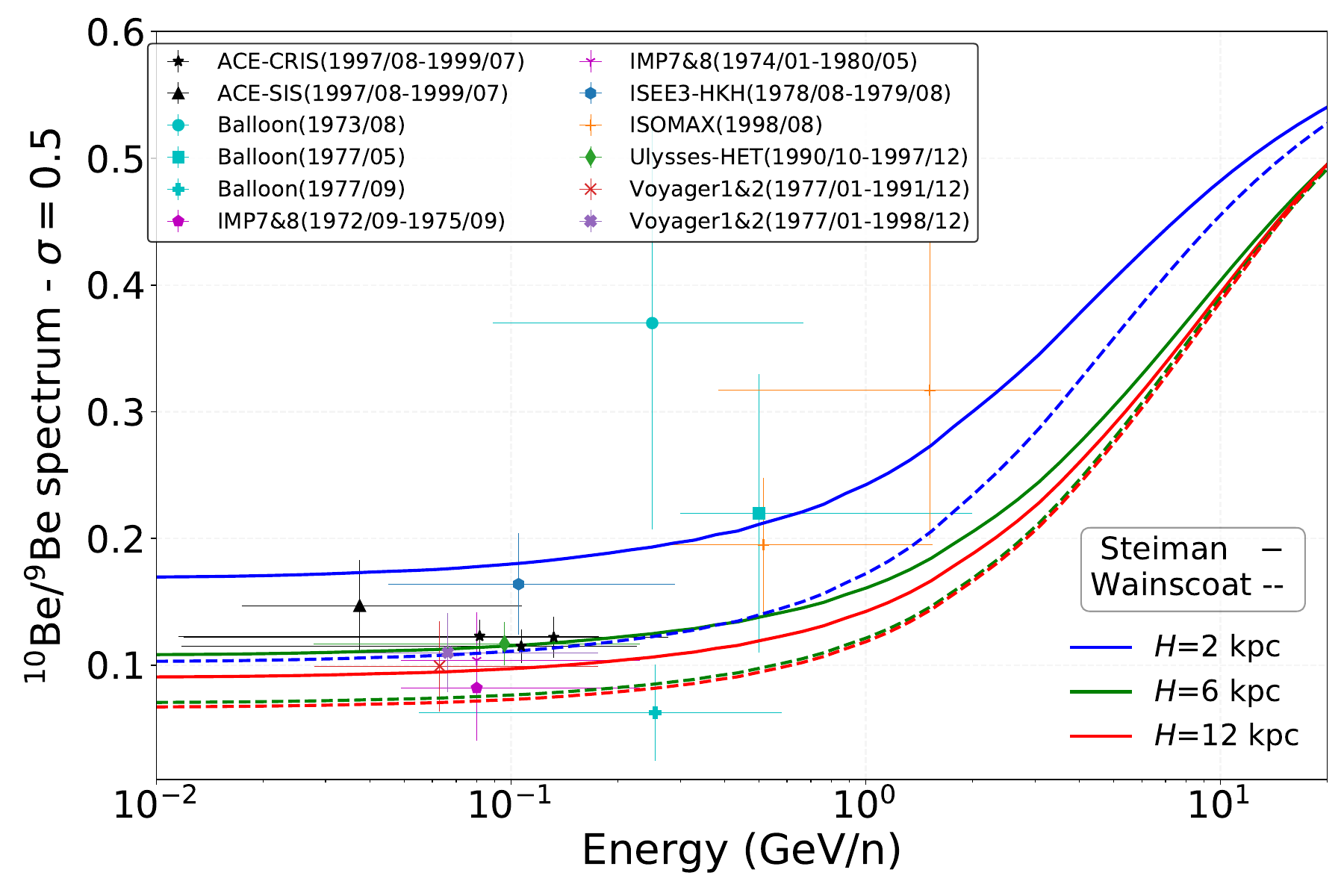}
\includegraphics[width=0.496\textwidth] {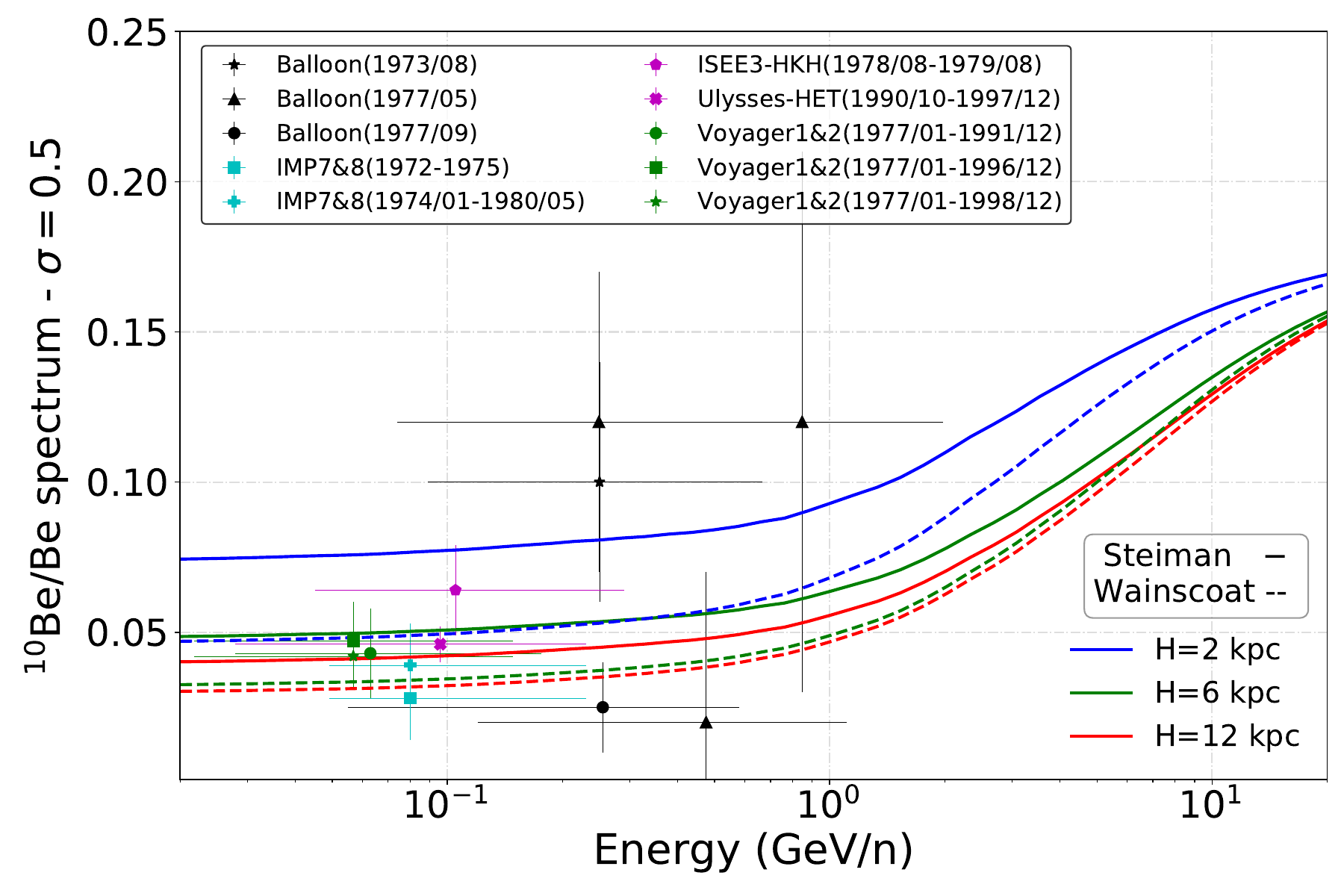}

\includegraphics[width=0.496\textwidth] {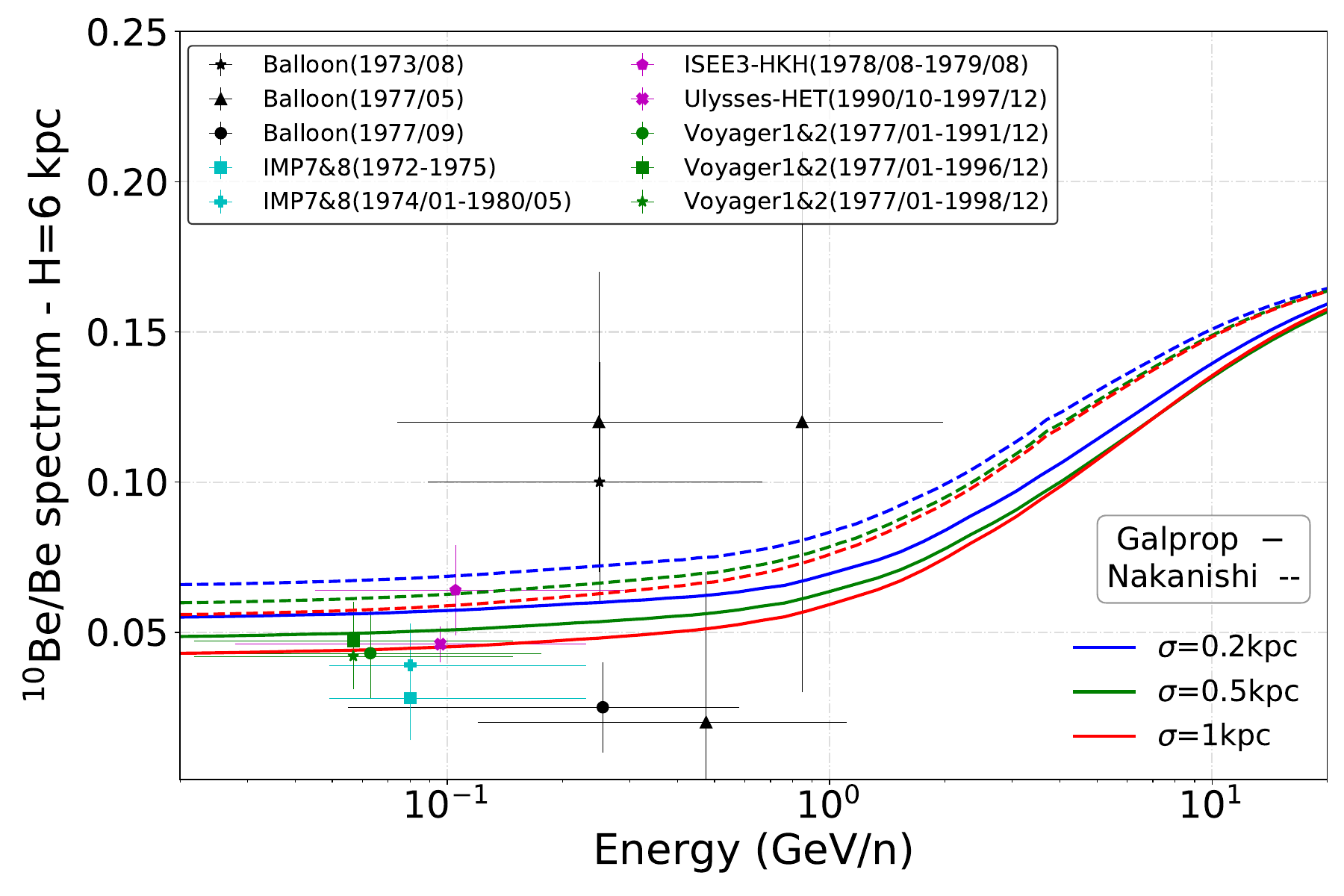} 
\includegraphics[width=0.496\textwidth] {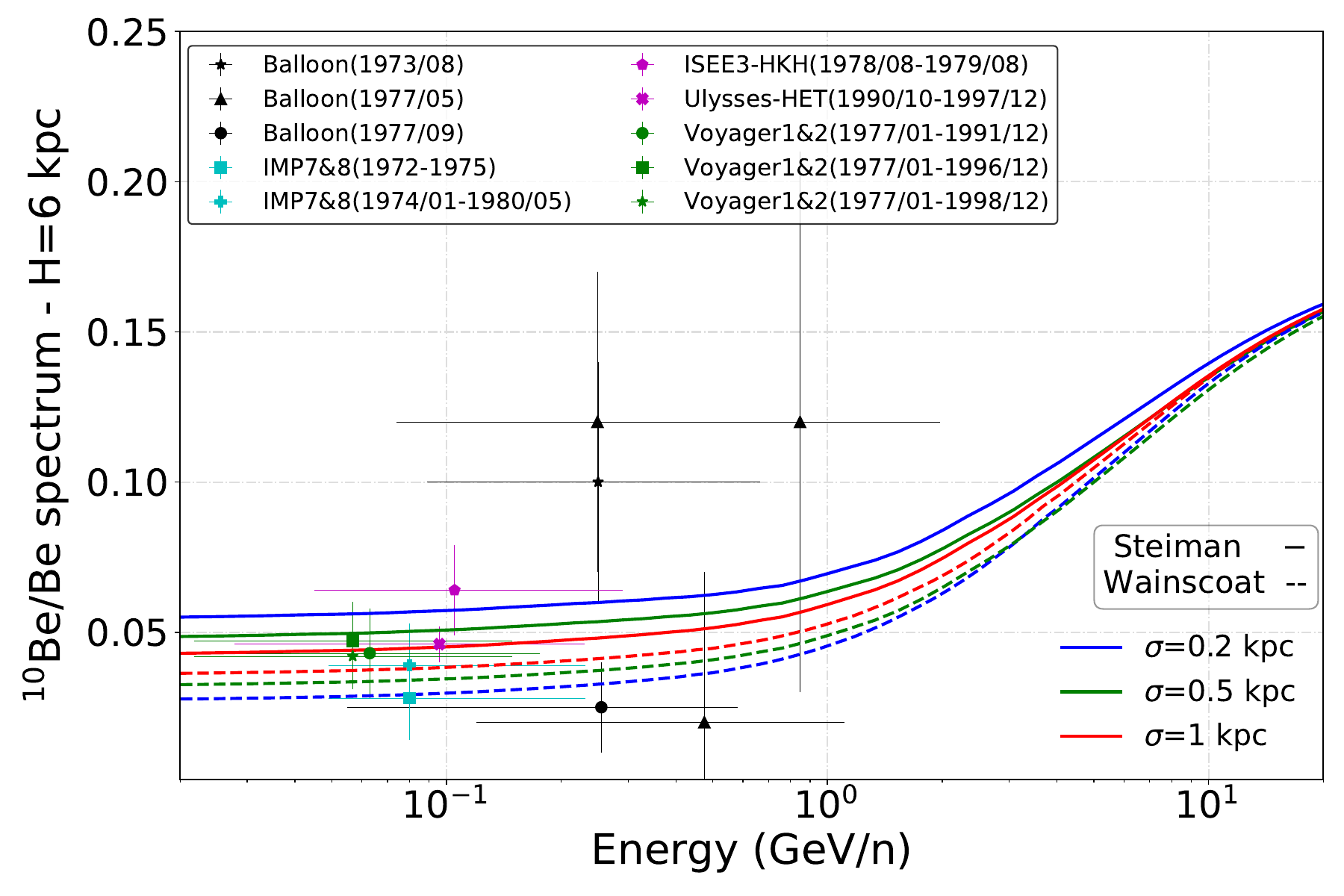} 
\caption{\textbf{Top panels:} Effect of changing the spiral arm model in the $^{10}$Be/$^{9}$Be (left) and $^{10}$Be/Be ratio (right) for different values of the halo height. \textbf{Bottom panels:} Same as in fig.~\ref{fig:Be10_Sigmas} but for the expected $^{10}$Be/Be local ratio.}
\label{fig:Be10_HSigma}
\end{figure}

In the top panels of Fig.~\ref{fig:Be10_HSigma}, we compare the $^{10}$Be/$^{9}$Be ratio (left) and $^{10}$Be/Be ratio (right) predicted using the Steiman-Cameron (solid lines) and Wainscoat (dashed lines) models for halo heights of $2$, $6$ and $12$~kpc. As discussed above, we see that the lower is the halo height, the larger is the difference in the predicted ratios. In both panels, we adopt a width of the spiral arm of $\sigma = 0.5$~kpc.
In the bottom panels, we show the same as in Fig.~\ref{fig:Be10_Sigmas} but for the expected $^{10}$Be/Be local ratio.

For completeness, in Table~\ref{tab:HComps} we provide the relative difference in the predicted $^{10}$Be/$^{9}$Be ratio for different variations of the gas model, at $10$, $100$ and $1000$~MeV and for values of the halo height of $2$, $6$ and $12$~kpc. In the left part of the table we report the relative difference of the predicted ratio in the benchmark 3D gas model (Steiman-Cameron 4-arm model) over the predicted ratio in the benchmark 2D gas model (Galprop model). In the right part of the table, we report the predicted relative differences of the ratio in the case of the benchmark 3D gas model over the Wainscoat spiral arm model.
\begin{table}[tb!]
\centering
\resizebox*{0.99\columnwidth}{0.13\textheight}{
\begin{tabular}{|c|c|c|c||c|c|c|}
\hline
  & \multicolumn{3}{|c||}{\textbf{\large{Galaxy structure (3D/2D)}}} &
\multicolumn{3}{|c|}{\textbf{\large{Spiral arm model (Steiman/Wains.)}}} \\
\hline & \textbf{10 MeV} & \textbf{100 MeV} & \textbf{1 GeV} & \textbf{10 MeV} & \textbf{100 MeV} & \textbf{1 GeV} \\ 
\hline
\multirow{1}{7em}{\centering \textbf{H$=2$~kpc}} & 1.341 & 1.331 & 1.25 & 1.645 & 1.62 & 1.407  \\ \hline 

\multirow{1}{7em}{\centering \textbf{H$=6$~kpc}} & 1.226 & 1.218 & 1.176 & 1.533 & 1.509 & 1.325  \\ \hline 

\multirow{1}{7em}{\centering \textbf{H$=12$~kpc}} &1.168 & 1.167 & 1.191 & 1.352 & 1.333 & 1.202 \\ \hline 
\end{tabular}
}
\caption{Relative difference of the predicted $^{10}$Be/$^{9}$Be ratio for different variations of the gas model, at $10$, $100$ and $1000$~MeV and for values of the halo height of $2$, $6$ and $12$~kpc, assuming a spiral arm width of $\sigma = 0.5$. In the left part, we report the predicted ratio in the benchmark 3D gas model (Steiman-Cameron 4-arm model) over the predicted ratio in the benchmark 2D gas model (Galprop model). In the right part, we report the predicted relative differences of the ratio obtained with the benchmark 3D gas model over that obatined with the Wainscoat spiral arm model.}
\label{tab:HComps}
\end{table}

Similarly,in Table~\ref{tab:SigmaComps} we report the relative differences in the predicted $^{10}$Be/$^{9}$Be ratio for different choices of the radial trend of the gas density (comparing the Galprop and Nakanishi models), in the left part of the table, and comparing the predicted ratios with the Steiman-Cameron and Wainscoat models, for different values of the arm width, from $\sigma=0.2$~kpc to $\sigma=1$~kpc.
\begin{table}[t!]
\centering
\resizebox*{0.99\columnwidth}{0.13\textheight}{
\begin{tabular}{|c|c|c|c||c|c|c|}
\hline
  & \multicolumn{3}{|c||}{\textbf{Gas model (GALPROP/Nikanishi)}} &
\multicolumn{3}{|c|}{\textbf{\large{Spiral arm model (Steiman/Wains.)}}} \\
\hline & \textbf{10 MeV} & \textbf{100 MeV} & \textbf{1 GeV}& \textbf{10 MeV} & \textbf{100 MeV} & \textbf{1 GeV}\\ 
\hline
\multirow{1}{7em}{\centering \textbf{$\sigma=0.2$~kpc}} & 0.844 & 0.839 & 0.831 &  2.072 & 2.001 & 1.576  \\ \hline 

\multirow{1}{7em}{\centering \textbf{$\sigma=0.5$~kpc}} & 0.819 & 0.814 & 0.803 &  1.533 & 1.509 & 1.325 \\ \hline 

\multirow{1}{7em}{\centering \textbf{$\sigma=1$~kpc}} & 0.785 & 0.78 & 0.782& 1.21 & 1.201 & 1.137 \\ \hline 
\end{tabular}
}
\caption{Similar to Table~\ref{tab:HComps} but comparing the ratios obtained for different arm width values, $\sigma$, assuming $H=6$~kpc. In the left part we compare the $^{10}$Be/$^{9}$Be ratio obtained for different choices of the radial trend of the gas density (Galprop and Nakanishi models), while in the right part we report the predicted ratio obtained with the Steiman-Cameron model over that obtained with the Wainscoat model.}
\label{tab:SigmaComps}
\end{table}

\begin{figure}[!t]
\centering
\includegraphics[width=0.496\textwidth] {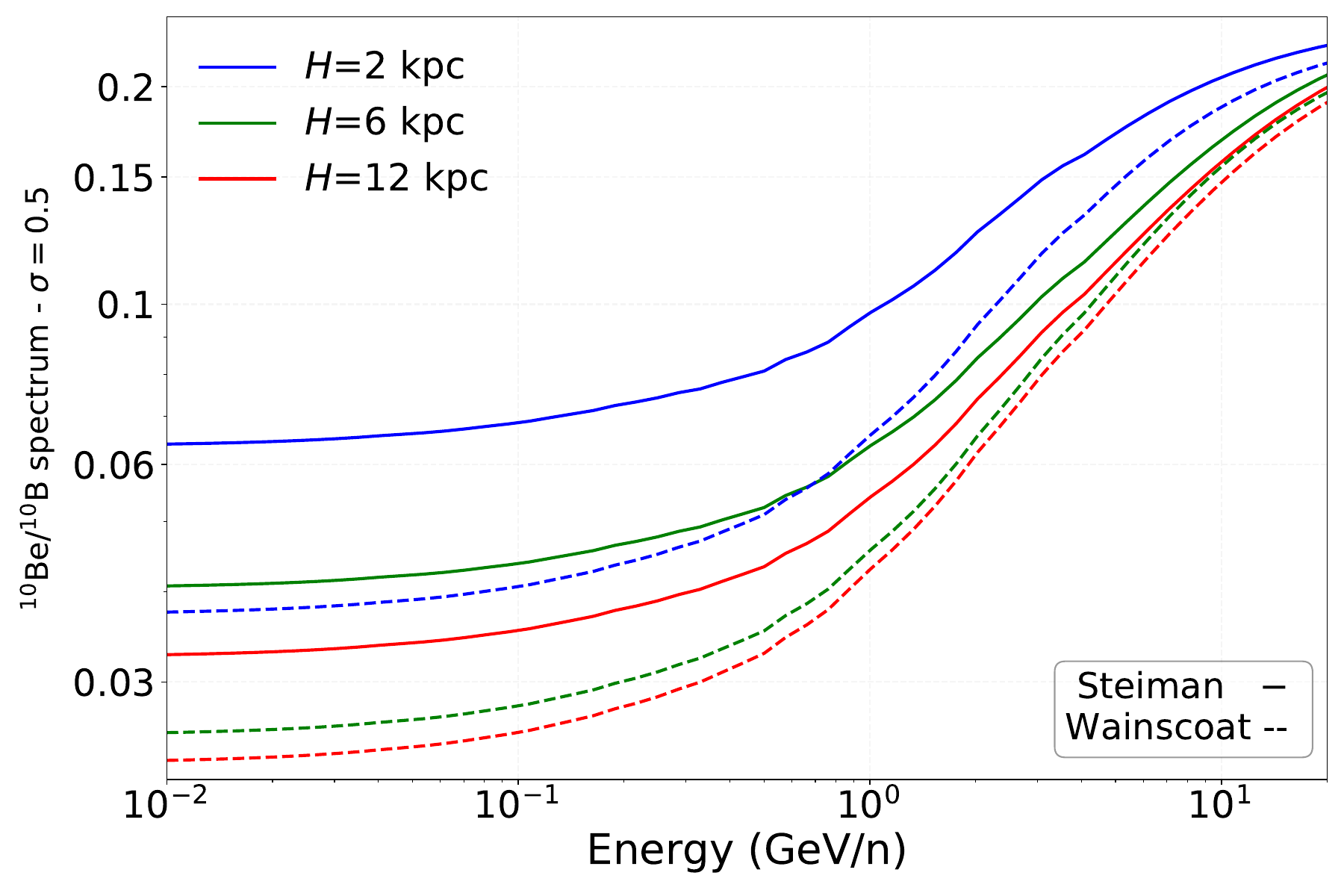}
\includegraphics[width=0.496\textwidth] {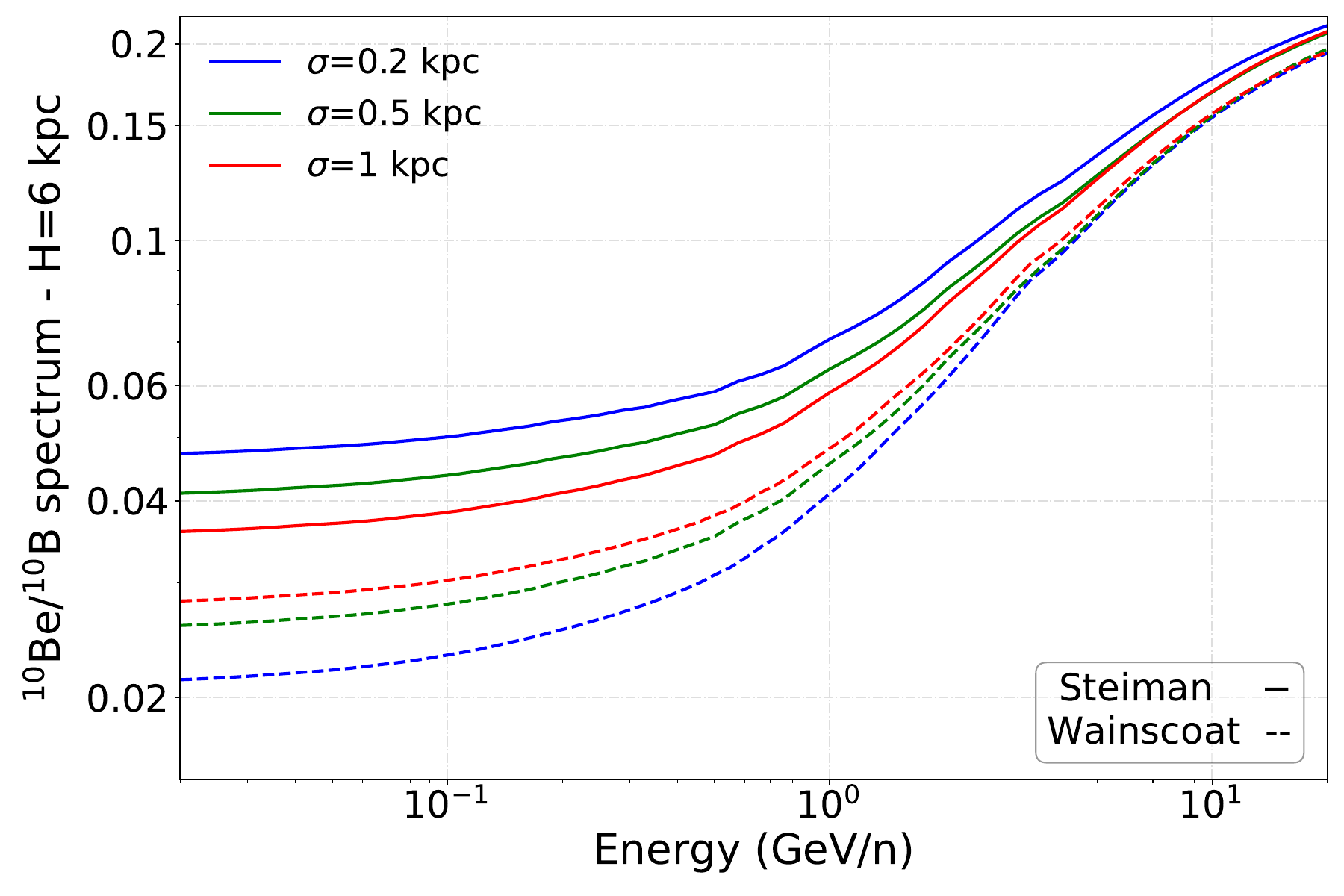} 
\includegraphics[width=0.495\textwidth] {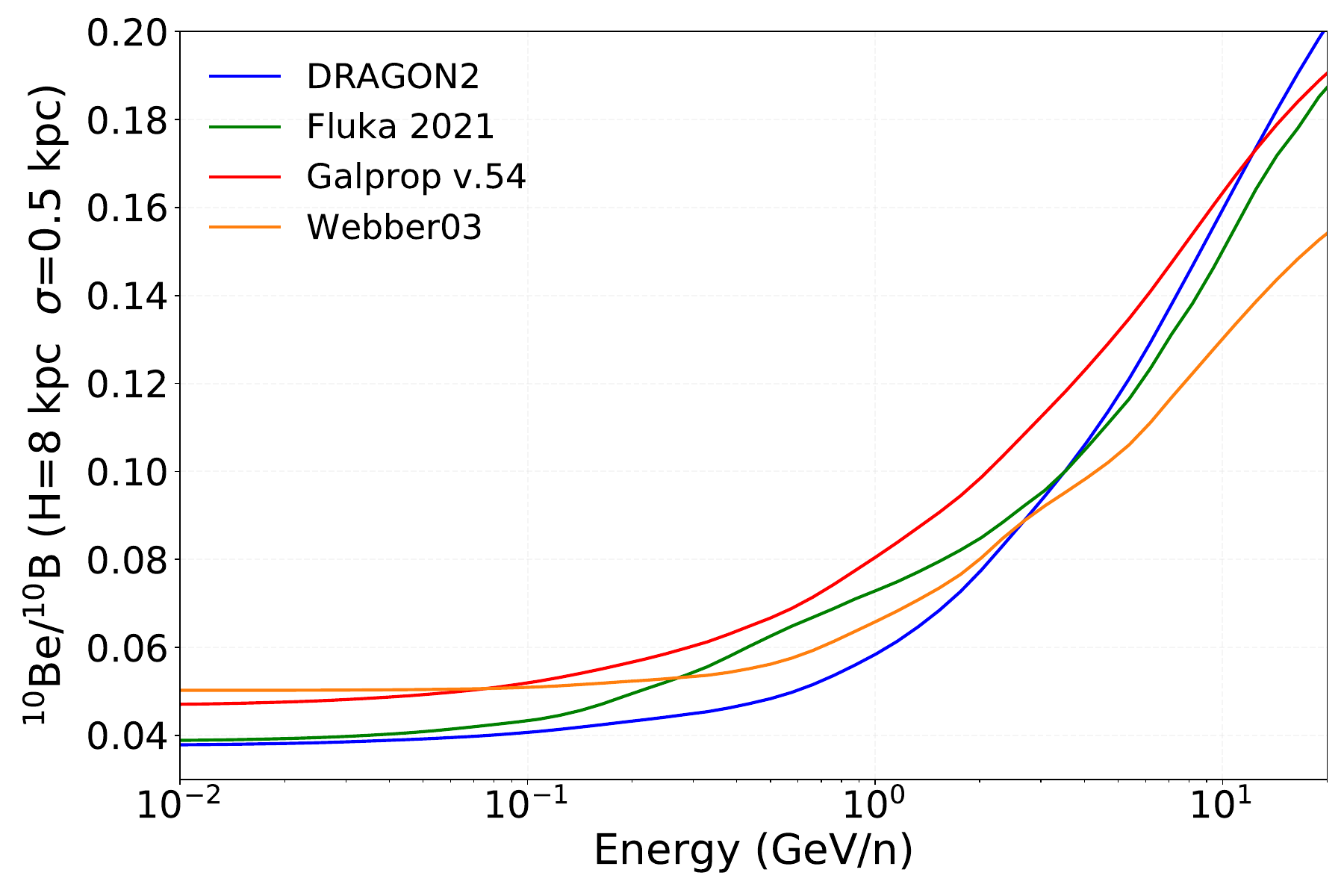}
\caption{The top panels show the predicted $^{10}$Be/$^{10}$B ratio for different choices of the gas density distribution, while the bottom panel shows the predicted ratio for different choices of cross sections parameterizations, assuming the Steiman-Cameron spiral arm model with $\sigma = 0.5$~kpc and $H=8$~kpc.}
\label{fig:Be10B10_ratios}
\end{figure}

Different authors have remarked on the importance of the Be/B ratio as a tool to determine the halo height, since it partially cancels other systematic uncertainties~\cite{CarmeloBeB}. The main reason is that $^{10}$Be decays into $^{10}$B, so that the ratio would be doubly affected by this decay and, therefore, the halo height. This effect is expected to be even more remarkable in the $^{10}$Be/$^{10}$B ratio, for which AMS-02 is expected to offer measurements soon, and for which HELIX will certainly be sensitive enough to detect. In Fig.~\ref{fig:Be10B10_ratios} we provide the predicted $^{10}$Be/$^{10}$B ratio for different choices of the gas density distribution in the top panels, and the predicted ratio using different cross sections parameterizations, in the bottom panel.

\begin{figure}[!t]
\centering
\includegraphics[width=0.6\textwidth] {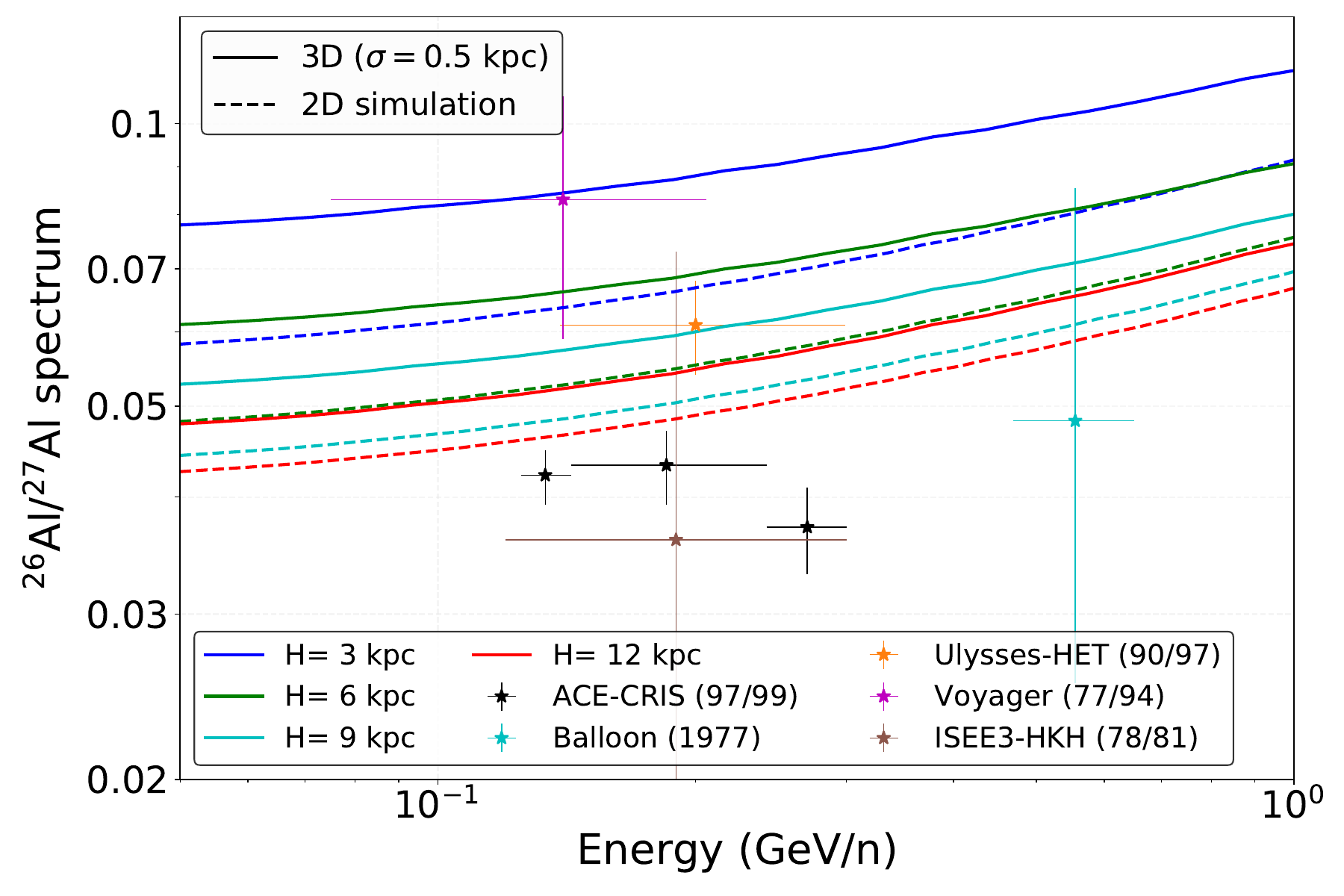} 

\includegraphics[width=0.495\textwidth] {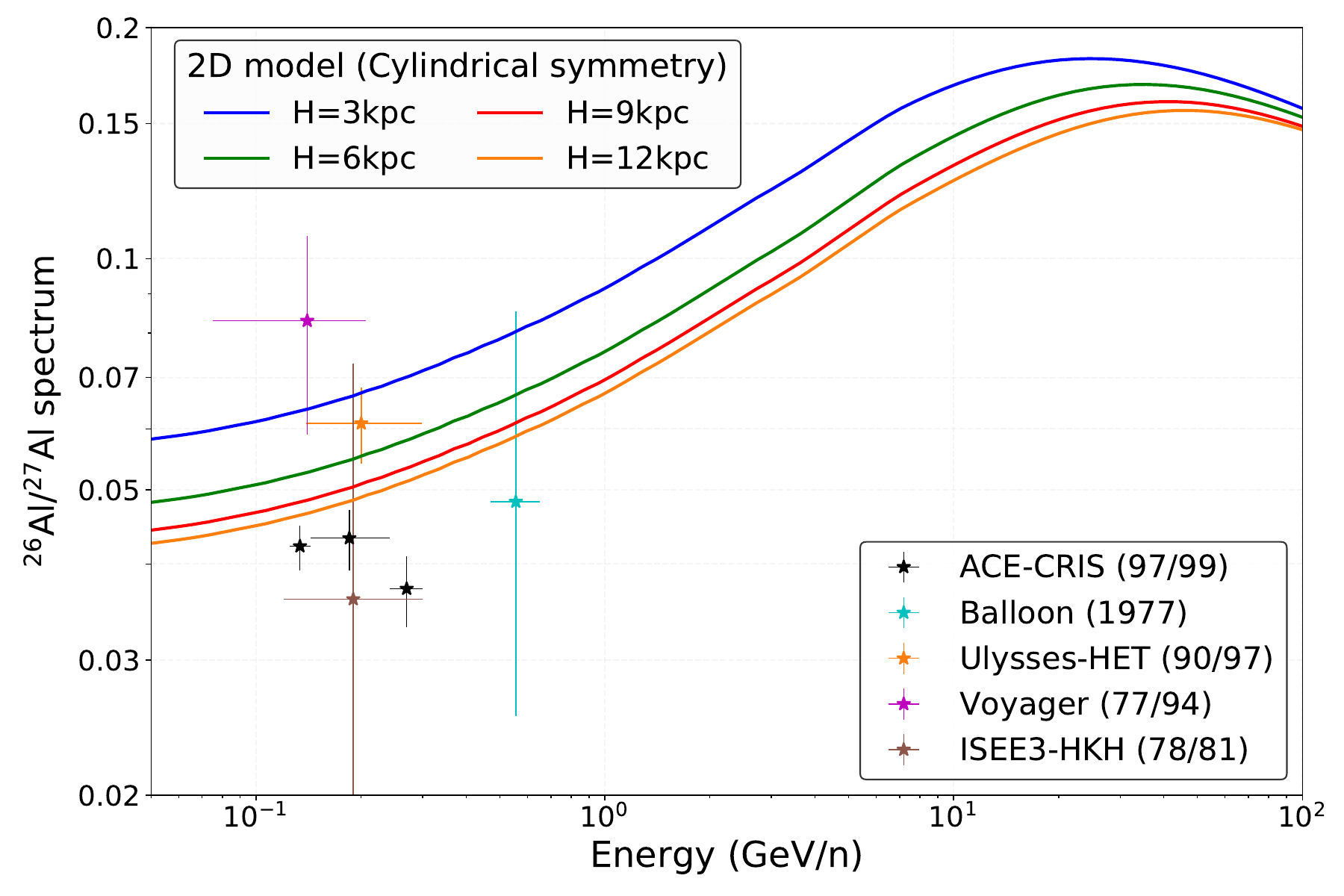} 
\includegraphics[width=0.495\textwidth] {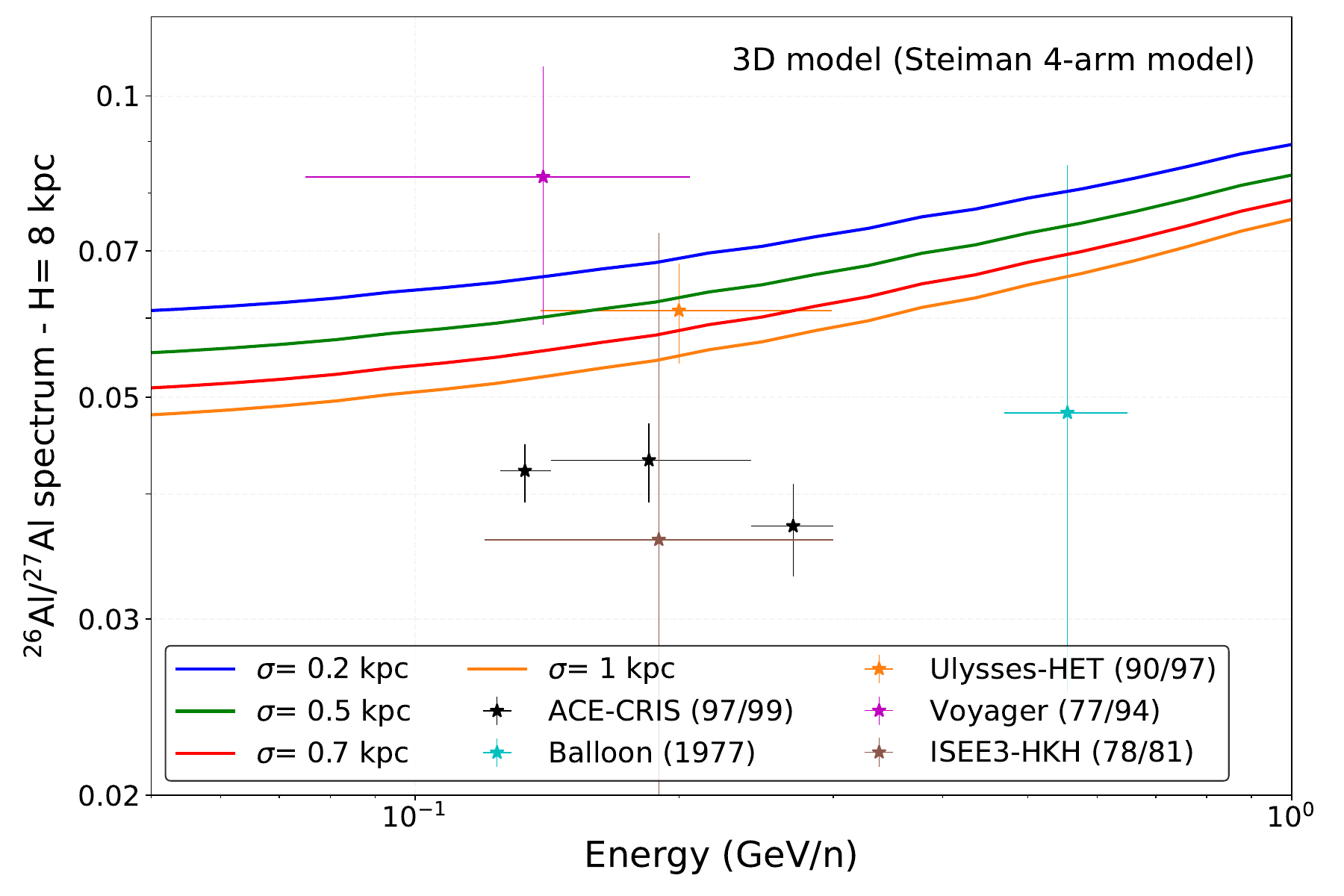}
\caption{\textbf{Top panel:} Similar to the left panel of Fig.~\ref{fig:Be10_2Dvs3D} but for the $^{26}$Al/$^{27}$Al ratio. \textbf{Bottom panels:} Predicted $^{26}$Al/$^{27}$Al ratio adopting the Galprop 2D gas model for different values of the halo height (left) and  adopting the Steiman-Cameron spiral arm model ($H=8$~kpc) for different values of the spiral arm width (right). 
}
\label{fig:Al_3D_2D}
\end{figure}

To illustrate that the gas distribution similarly affects the local fluxes of other unstable CR nuclei (with lifetimes around 1~Myr), in Fig.~\ref{fig:Al_3D_2D} we show the predicted ratios of $^{26}$Al/$^{27}$Al for our benchmark 2D and 3D gas models.
$^{26}$Al is an unstable isotope that decays into $^{26}$Mg through $\beta^+$ decay with a half-life of $\sim0.72$~Myr~\cite{1983LPSC...14..331N} and also via electron capture, with a half-life of $\sim6.3$~s. Therefore, in DRAGON2 we implement the propagation of $^{26}$Al having into account the probability of capturing an electron from the ISM (which depends on the gas model used, although different distributions no not alter significantly the predicted local flux of $^{26}$Al) and its subsequent decay.

Although challenging, the detection of this isotope from AMS-02 can be reached within the next $10$~yr.
We emphasize that in our calculation, $^{26}$Al is treated as a pure secondary CR, not including any primary injection. In fact, this isotope is expected to be significantly produced in supernovae, and measurements of the diffuse gamma-ray line at $1.8$~MeV seem to support supernovae as the dominant production channel~\cite{Pleintinger_2023}, but only for energies below $\sim$ a few MeV. Therefore, it is expected that above tens of MeV, $^{26}$Al is predominantly produced in CR interactions and can explain the current measurements, also in agreement with previous estimations (see .e.g. Fig.~5 of Ref.~\cite{Boschini_2022}).
As we see from the top panel of Figure~\ref{fig:Al_3D_2D}, the effect of adopting the spiral arm pattern of the gas distribution is similar to the case of $^{10}$Be, as expected. Other heavy unstable nuclei have been used in the past as CR clocks, and will suffer from similar uncertainties from the gas distributions as in the case of $^{10}$Be or $^{26}$Al.

For clearness, we show in the left panel of Fig.~\ref{fig:Al_3D_2D} the predicted ratio for different values of the halo height and under the Galprop 2D gas model, in the energy range between $0.05$~GeV/n to $100$~GeV/n. In the right panel, we display the predicted ratio using the Steiman-Cameron spiral arm model ($H=8$~kpc) for different values of the width of the spiral arm. 

\newpage

\section{Fit to spectra of primary CRs and secondary ratios}
\label{sec:AppendixC}

\begin{figure}[!t]
\centering
\includegraphics[width=0.441\textwidth] {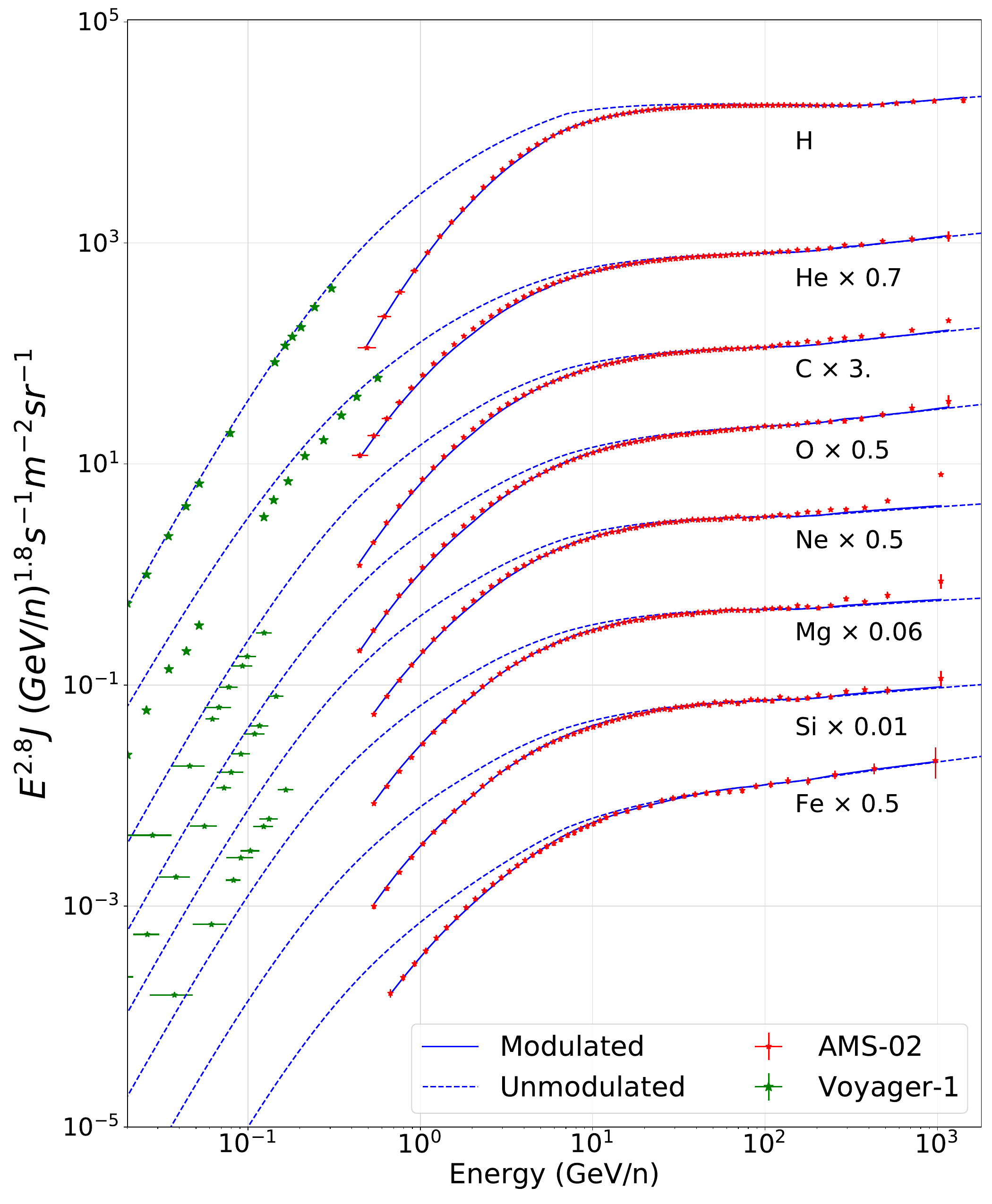}
\caption{Comparison of the primary CR spectra with AMS-02 and Voyager-1 data. adopting the Steiman-Cameron spiral arm model with a halo height of $8$~kpc and an arm width of $0.5$~kpc}
\label{fig:Prims}
\end{figure}

In this section, we compare the predicted diffuse flux and CR ratios with experimental data from AMS-02. In Fig.~\ref{fig:Prims} we show a comparison between the primary CR spectra that are fit to a combination of AMS-02~\cite{Aguilar:2015ooa, Aguilar:2020ohx, Aguilar:2021tla, Aguilar:2018keu} and Voyager-1 data~\cite{cummings2016galactic, stone2013voyager}. The injection parameters are adjusted following the work in Ref.~\cite{Luque_MCMC}. These parameters can be seen directly in the input DRAGON files that are available at \url{https://github.com/tospines/Codes-Models-ExtraPlots/tree/main/10Be_GasDistribution}.
In each of these files, one can also find the parameters that model the spiral arm distributions in each of our models, which are parameterized as described in App.C.9 of Ref.~\cite{DRAGON2-1}. 
We have checked the consistency of our results for DRAGON spatial resolutions down to $\sim100$~pc, and adopted a reference spatial resolution of $200$~pc in our default studies. 

\begin{table}[b!]
\centering
\resizebox*{0.6\columnwidth}{0.04\textheight}{
\begin{tabular}{|c|c|c|c|c|}
\hline \textbf{D$_0$~($\times10^{28}$ cm$^2$/s)} & \textbf{R$_0$ (GV)} & \textbf{$\eta$} & \textbf{V$_A$ (km/s)} & \textbf{$\delta$} \\ 
\hline
\multirow{1}{5em} {\centering \, 10.2 \,\,} & 4 &\, -0.75 \,& 13.43 & \, 0.49 \, \\ \hline 
\end{tabular}
}
\caption{Propagation parameters for simulations that adopt the Steiman-Cameron spiral arm model for a halo height of $8$~kpc and $\sigma=0.5$~kpc, with DRAGON2 cross sections.}
\label{tab:PropParams}
\end{table}

\begin{figure}[!h]
\centering
\includegraphics[width=0.45\textwidth] {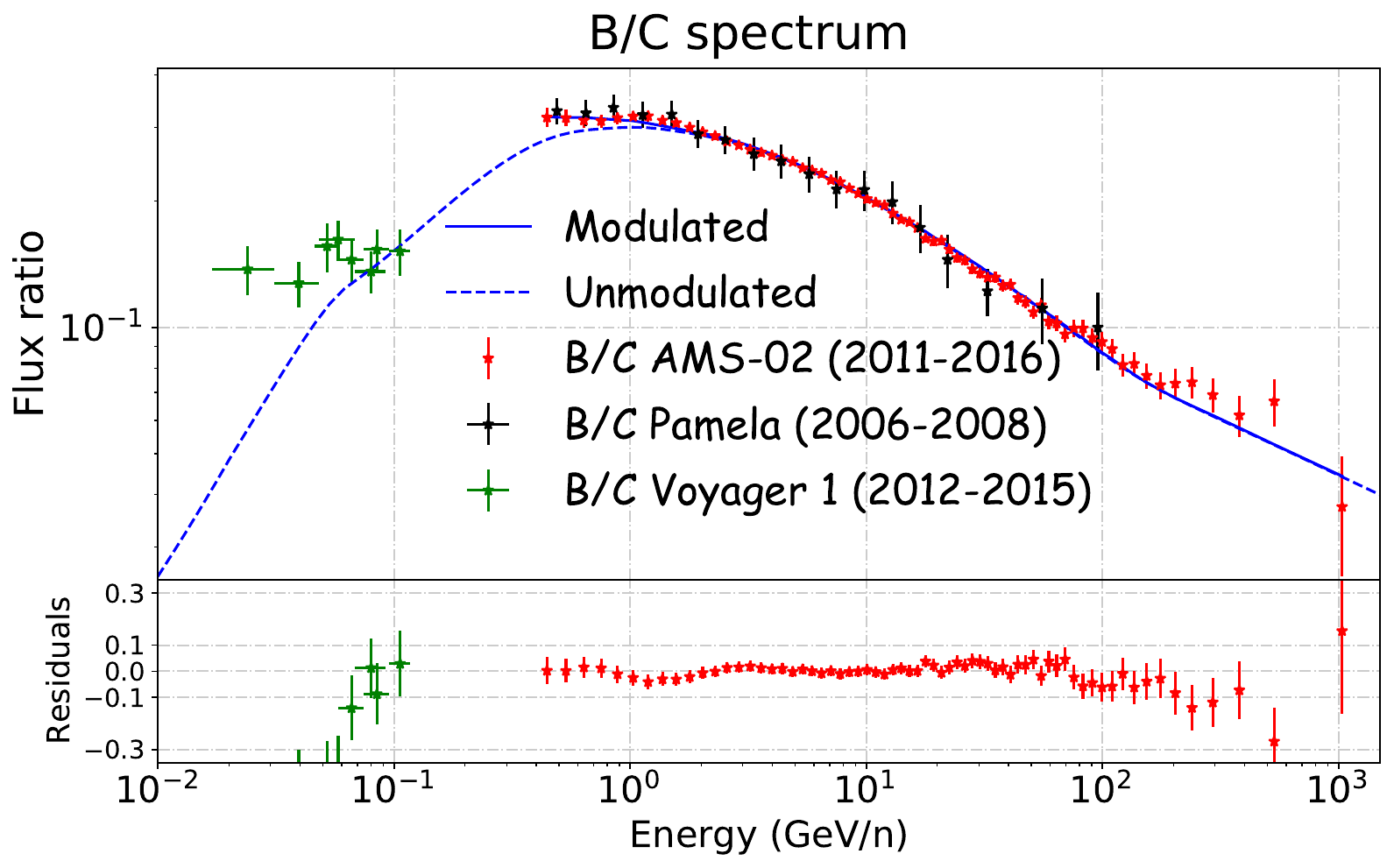}

\includegraphics[width=0.44\textwidth] {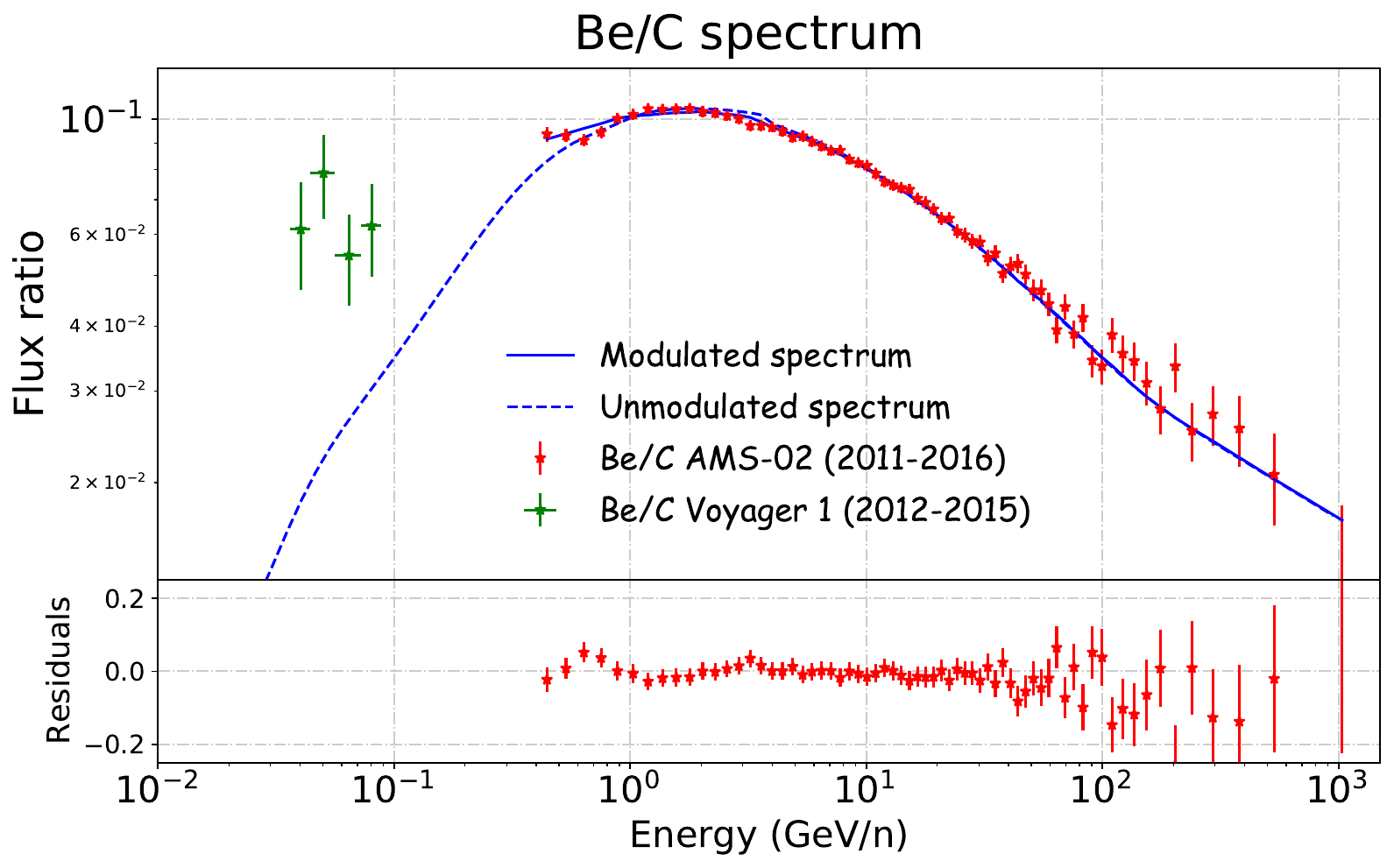} 
\includegraphics[width=0.44\textwidth] {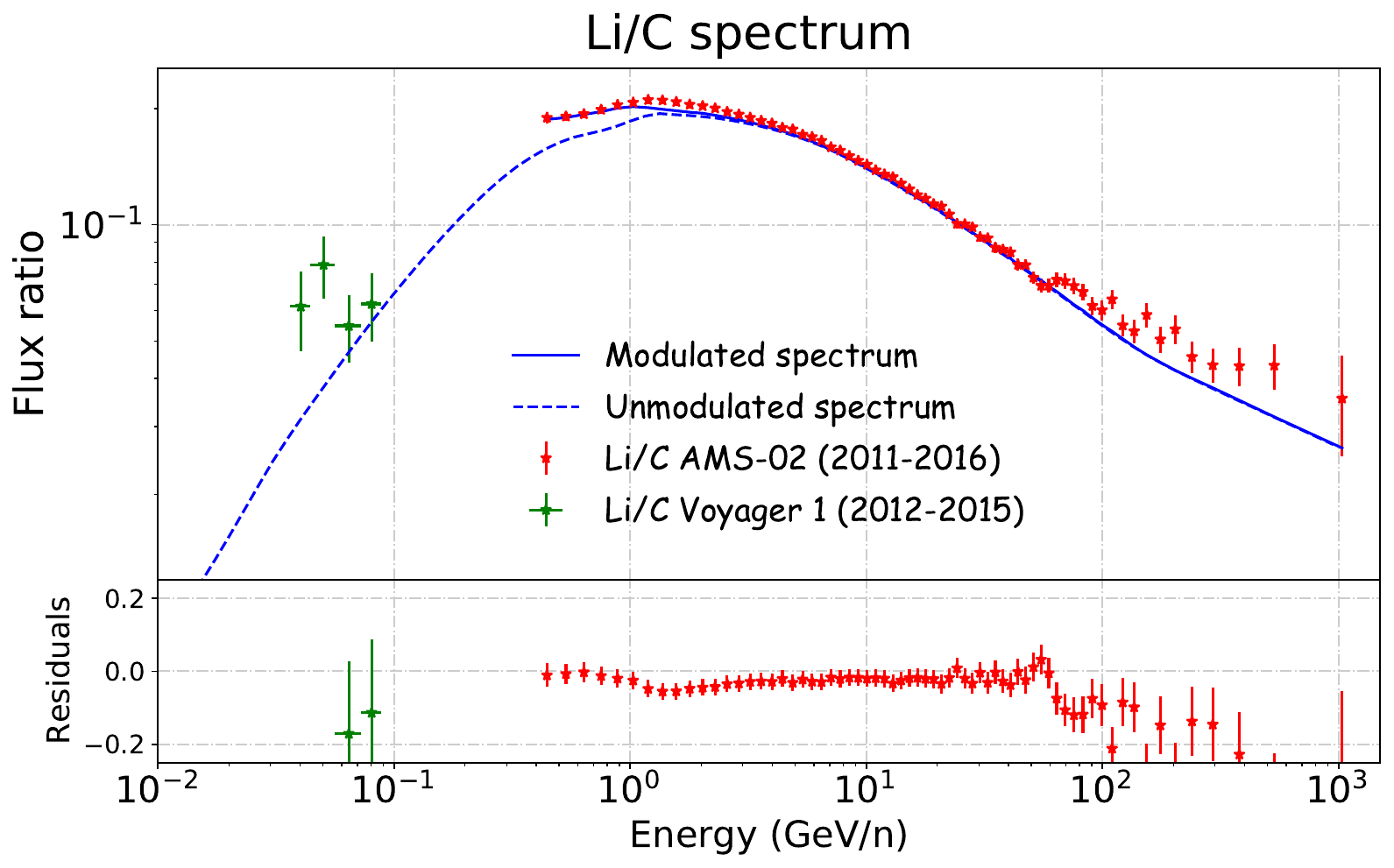}
\caption{Similar to Fig.~\ref{fig:Prims} but for the secondary-to-primary ratios of B, Be and Li to carbon.}
\label{fig:Secs}
\end{figure}

Similarly, we show the fit to secondary-to-primary CR ratios~\cite{aguilar2017observation} in Fig.~\ref{fig:Secs}. In both figures, we report the results using the Steiman-Cameron spiral arm model for sources and gas distribution, adopting a halo height of $8$~kpc and an arm width of $0.5$~kpc. We remind the reader that our fits do not incorporate correlations in AMS-02 errors.
As a reference, the propagation parameters in this case are those reported in Table~\ref{tab:PropParams}.

\end{document}